\documentclass[acmsmall]{acmart}

\usepackage{multirow}
\usepackage{enumitem}

\usepackage{subfigure}
\usepackage{xspace}
\usepackage{multirow}
\usepackage{xcolor}
\usepackage{amsmath}
\usepackage{amsfonts}
\usepackage{bm}
\usepackage{soul}
\usepackage[ruled]{algorithm2e}
\usepackage{balance}
\usepackage[most]{tcolorbox}
\usepackage{longtable}
\usepackage{makecell}
\usepackage{url}
\usepackage{hyperref}
\usepackage{booktabs}
\usepackage{tabularray}
\usepackage{ltablex}
\keepXColumns

\AtBeginDocument{%
  \providecommand\BibTeX{{%
    \normalfont B\kern-0.5em{\scshape i\kern-0.25em b}\kern-0.8em\TeX}}}

\acmJournal{TIST}

\newcommand{\rev}[1]{{\color{black}{#1}}}

\newcommand{\eat}[1]{}
\newcommand{\zhaobite}[1]{}
\newcommand{\subsec}[1]{{\noindent- \textbf{\textit{#1}}}}

\newcommand{\eg}{\emph{e.g.},\xspace}
\newcommand{\ie}{\emph{i.e.},\xspace}

\newcommand{\etc}{\emph{etc.}\xspace}

\newcommand\figref[1]{Figure~\ref{#1}}

\newcommand\secref[1]{Section~\ref{#1}}

\newtheorem{defi}{\textbf{Definition}}

\begin{document}

\title{Towards Urban General Intelligence: A Review and Outlook of Urban Foundation Models}






\author{Weijia Zhang}
\affiliation{%
  \institution{The Hong Kong University of Science and Technology (Guangzhou)}
  \city{Guangzhou}
  \country{China}
}
\email{wzhang411@connect.hkust-gz.edu.cn}

\author{Jindong Han}
\affiliation{%
  \institution{Shandong University}
  \country{China}
}
\email{jindong.han@sdu.edu.cn}

\author{Zhao Xu}
\affiliation{%
  \institution{The Hong Kong University of Science and Technology (Guangzhou)}
  \city{Guangzhou}
  \country{China}
}
\email{zxu674@connect.hkust-gz.edu.cn}

\author{Hang Ni}
\affiliation{%
  \institution{The Hong Kong University of Science and Technology (Guangzhou)}
  \city{Guangzhou}
  \country{China}
}
\email{hni017@connect.hkust-gz.edu.cn}

\author{Tengfei~Lyu}
\affiliation{%
  \institution{The Hong Kong University of Science and Technology (Guangzhou)}
  \city{Guangzhou}
  \country{China}
}
\email{tlyu077@connect.hkust-gz.edu.cn}

\author{Hao Liu}
\authornote{Corresponding author.}
\affiliation{%
  \institution{The Hong Kong University of Science and Technology (Guangzhou) and The Hong Kong University of Science and Technology}
  \country{China}
}
\email{liuh@ust.hk}

\author{Hui Xiong}
\affiliation{%
  \institution{The Hong Kong University of Science and Technology (Guangzhou) and The Hong Kong University of Science and Technology}
  \country{China}
}
\email{xionghui@ust.hk}


\begin{abstract}
The integration of machine learning techniques has become a cornerstone in the development of intelligent urban services, significantly contributing to the enhancement of urban efficiency, sustainability, and overall livability. Recent advancements in foundational models, such as ChatGPT, have introduced a paradigm shift within the fields of machine learning and artificial intelligence. These models, with their exceptional capacity for contextual comprehension, problem-solving, and task adaptability, present a transformative opportunity to reshape the future of smart cities and drive progress toward Urban General Intelligence~(UGI).
Despite increasing attention to Urban Foundation Models (UFMs), this rapidly evolving field faces critical challenges, including the lack of clear definitions, systematic reviews, and universalizable solutions. 
To address these issues, this paper first introduces the definition and concept of UFMs and highlights the distinctive challenges involved in their development.
Furthermore, we present a data-centric taxonomy that classifies existing research on UFMs according to the various urban data modalities and types. 
In addition, we propose a prospective framework designed to facilitate the realization of versatile UFMs, aimed at overcoming the identified challenges and driving further progress in this field. 
\rev{Finally, this paper systematically summarizes and discusses existing benchmarks and datasets related to UFMs, and explores the wide-ranging applications of UFMs within urban contexts, illustrating their potential to significantly impact and transform urban systems.}
A comprehensive collection of relevant research papers and open-source resources have been collated and
are continuously updated at: \href{https://github.com/usail-hkust/Awesome-Urban-Foundation-Models}{https://github.com/usail-hkust/Awesome-Urban-Foundation-Models}.
\end{abstract}

\ccsdesc[500]{Information systems~Spatial-temporal systems}
\ccsdesc[500]{Computing methodologies~Machine learning}


\maketitle

\section{Introduction}

\begin{tcolorbox}[breakable, colback=gray!5!white, colframe=gray]
    \textit{Urban General Intelligence (UGI) refers to a conceptualized advanced form of artificial intelligence tailored to understand, interpret, and adeptly manage complex urban systems and environments. Analogous to Artificial General Intelligence (AGI),  UGI is envisioned to autonomously perform any intellectual task related to urban contexts, rivaling or even surpassing human capabilities, thereby transforming cities into more livable, resilient, and adaptive spaces.}
\end{tcolorbox} 


Machine learning technologies have become integral to reshaping urban environments, serving as the foundation for various smart city initiatives. By enabling more effective resource distribution, enhancing public services, and improving residents' quality of life, these technologies significantly contribute to urban intelligence~\cite{ullah2020applications}.
Leveraging extensive datasets from sources such as urban sensors, satellite imagery, and social media platforms, machine learning algorithms uncover complex urban patterns and provide precise forecasts of city dynamics~\cite{han2024bigst,zhang2020semi}.
This analytical capability is critical for optimizing urban operations, such as intelligent energy management, traffic flow regulation, and environmental surveillance, thereby advancing the development of smarter, more sustainable cities~\cite{nosratabadi2019state, kyrkou2022machine}.


The emergence of foundation models, such as Large Language Models (LLMs, \eg~ChatGPT) and Vision Foundation Models (VFMs), has profoundly transformed the research paradigm in machine learning and artificial intelligence~\cite{zhou2023comprehensive}.
These models, distinguished by their extensive pre-training on massive datasets, exhibit remarkable emergent capabilities, such as contextual reasoning, complex problem-solving, and zero-shot adaptability across a wide range of tasks~\cite{bommasani2021opportunities,kaplan2020scaling}.
Such capabilities position foundation models as ideal tools for engaging with the complexities of dynamic and multifaceted urban environments by seamlessly integrating diverse data sources, enabling real-time and versatile analysis, and facilitating adaptive decision-making. This lays the groundwork for urban systems that are seamlessly integrated, highly intelligent, and adept at responding to evolving challenges.

\begin{figure*}[tb]
\centering
\includegraphics[width=1\columnwidth]{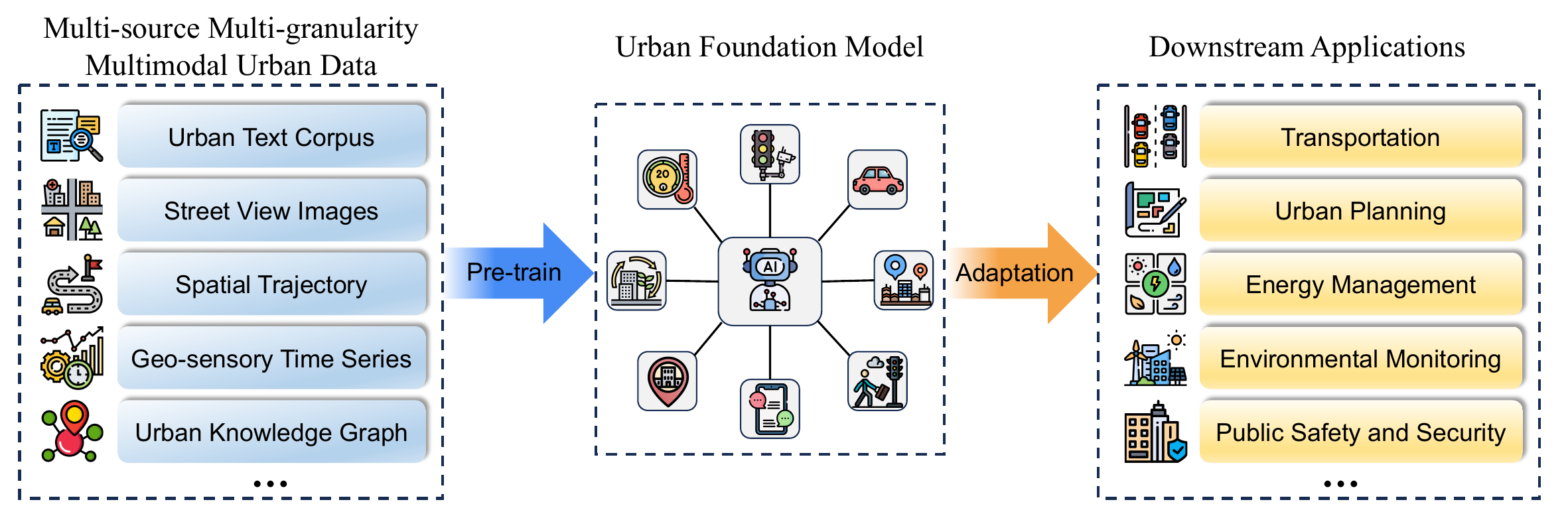}
\caption{Urban Foundation Models (UFMs) are pre-trained on multi-source, multi-granularity and multimodal urban data and can be adapted for a variety of downstream urban application domains.}
\label{fig:basic}
\end{figure*}

Urban Foundation Models (UFMs), as illustrated in \figref{fig:basic}, represent an emerging class of models that are pre-trained on a vast array of urban data sources, which are of various levels of granularity and multiple modalities. These models demonstrate a profound comprehension of diverse urban data features and exhibit remarkable adaptability to a wide range of urban applications~\cite{balsebre2023city}, playing a pivotal role in advancing the realization of Urban General Intelligence (UGI). By integrating and interpreting heterogeneous urban data, UFMs provide comprehensive insights, reveal complex spatiotemporal patterns, and significantly enhance decision-making processes across various urban challenges. For example, UFMs are capable of analyzing a combination of human instructions, visual scene data, traffic sensor readings, and GPS trajectories to optimize traffic efficiency and safety~\cite{zhang2023trafficgpt,zheng2023chatgpt,lai2023large,da2024open}. Moreover, UFMs can collectively leverage demographic, land use, and environmental data to generate critical insights for the sustainable development of urban areas~\cite{magee2023steamlining,wang2023towards,zhou2024large}.

While the burgeoning interest and potential of UFMs, this field faces significant challenges, including the absence of clear definitions, the lack of systematic reviews of existing literature, and the need for universally applicable solutions~\cite{xu2023urban, balsebre2023city, zhang2023trafficgpt, jakubik2023foundation, xie2023geo, mai2023opportunities}. 
To address these critical gaps, this paper delivers a comprehensive survey on UFMs. It begins by offering a clear definition and basic concepts of UFMs and outlining the distinctive challenges associated with their development and application. 
Building upon this, we present a data-centric taxonomy to categorize and encapsulate existing research in the field. This taxonomy, grounded in the diverse modalities and types of urban data, serves to illuminate the progress and concerted efforts made in this emerging field.
Furthermore, we propose a forward-looking framework for constructing versatile UFMs. This framework is strategically developed to address the identified challenges while having the potential for a broad generalization across a variety of urban tasks and domains. Finally, we explore the practical applications of UFMs, emphasizing their role in advancing various urban domains and enhancing diverse facets of urban intelligence.


\rev{This paper is a significantly expanded version of our previously accepted work~\cite{zhang2024urban} at the KDD'24 tutorial track (8 pages, accessible at https://dl.acm.org/doi/abs/10.1145/3637528.3671453).
Compared to the concise conference version, this extended paper makes the following major contributions and extensions:
(1)~This work formally defines and conceptualizes UGI, shedding light on transformative opportunities for achieving UGI through UFMs.
(2)~We substantially extend and refine the review and data-centric taxonomy on UFMs literature: a new section to introduce current key techniques of building UFMs; a new section to summarize existing benchmarks and datasets regarding UFMs; incorporating a dedicated section on geovector-based UFMs; adding several new subsections, such as introducing pre-training methods on ordinary time series, prompt engineering and model reprogramming methods for time series cross-modal adaptation, model fine-tuning and reprogramming methods for trajectory cross-modal adaptation; integrating the latest advancements and literature related to UFMs, with 65\% of the references being newly added, delivering a more comprehensive, up-to-date, and insightful review framework.
(3)~We propose a prospective general framework for the development of versatile UFMs, providing potential solutions to address the identified challenges and empower generalizability across diverse urban applications and dynamic environments.
}


\rev{The rest of this paper is organized as follows. In \secref{basic}, we introduce basic definitions and concepts of UFMs.  
\secref{challenge} highlights the unique challenges involved in building UFMs. 
\secref{overview} delves into existing studies on UFMs, categorized into language-based models, vision-based models, time series-based models, trajectory-based models, geovector-based models, multimodal models, and other models. \secref{solution} presents a prospective framework for building versatile UFMs. 
\secref{dataset} summarizes existing benchmarks and datasets related to UFMs.
\secref{application} discusses UFMs' promising applications in different urban domains. Finally, we conclude this paper in \secref{conclusion}.
}

\section{Basics of Urban Foundation Models} \label{basic}




\begin{defi}[\textbf{Urban Foundation Models}]
Urban Foundation Models (UFMs) are a family of large-scale models pre-trained on vast amounts of multi-source, multi-granularity, and multimodal urban data. 
They acquire notable general-purpose capabilities in the pre-training phase, exhibiting remarkable emergent abilities and adaptability dedicated to a range of urban application domains, such as transportation, urban planning, energy management, environmental monitoring, and public safety and security. 
\end{defi}

\subsection{Data Characteristics}
UFMs are distinguished by their ability to process and analyze a vast array of data types, each contributing uniquely to the comprehensive understanding of urban dynamics. The data characteristics of UFMs, including multi-source, multi-granularity, and multimodal aspects, are essential for capturing the complexities of urban environments. 

\subsec{Multi-source.} UFMs integrate large-scale data from diverse urban sources, including sensor networks, social media, satellite imagery, mobile devices, \etc, facilitating a comprehensive understanding of urban environments.

\subsec{Multi-granularity.} UFMs handle data at different levels of spatial and temporal granularity. Spatially, they analyze city-wide patterns like traffic flow and population movement at the macro level, while modeling localized behaviors in neighborhoods or individuals at the micro level. Temporally, the collected data can span from annual records to real-time measurements.

\subsec{Multimodal.} UFMs are required to integrate diverse data modalities, such as textual, visual, and sensor-based inputs, enabling them to capture urban complexities more effectively than unimodal models.

\subsection{Key Techniques of Building Urban Foundation Models}
Like foundation models in other fields, the construction of UFMs primarily relies on two type of techniques, \ie pre-training and adaptation.

\subsubsection{Pre-training}
UFMs typically undergo pre-training on extensive and diverse datasets including geo-text data, social media content, street view images, trajectories, spatiotemporal time series, Points of Interests~(POIs), road networks, \etc 
The primary goal of pre-training is to enable these models to obtain as much general knowledge and patterns as possible, thereby capturing the overarching characteristics and structures inherent in the data. Pre-training methods for UFMs are principally divided into four categories: supervised pre-training, generative pre-training, contrastive pre-training, and hybrid pre-training \cite{sslsurvey2021}:

\subsec{Supervised pre-training.} This method involves training models on datasets with a substantial volume of labeled data, pairing an input with its corresponding output \cite{dosovitskiy2020image, devlin2018bert}. It allows models to learn the relationship between inputs and outputs in a supervised manner. Suppose $\mathcal{D} = \{ (x_1, y_1), (x_2, y_2), \ldots, (x_N, y_N) \}$ is the training dataset, where $x_i$ represents the $i$-th sample and $y_i$ is the corresponding label. Here is an example of the most commonly used cross-entropy loss function~\cite{goodfellow2016deep}:
\begin{equation}
    L_{sup} = -\sum_{i=1}^{N} \sum_{c=1}^{C} y_{ic} \log(\hat{y}_{ic}) ,\label{l_sup}
\end{equation}
where $N$ is the number of samples, $C$ is the number of classes,  $y_{ic}$ is a binary indicator (0 or 1) to indicate if class label $c$ is the correct classification for observation $i$, $\hat{y}_{ic}$ is the predicted probability observation $i$ is of class $c$.

\rev{
\subsec{Generative pre-training.} This approach trains models to generate or reconstruct (masked) input data in the absence of labeled data~\cite{radford2018improving, radford2019language}. It necessitates the model's ability to autonomously discern patterns and regularities, thereby producing outputs that closely mimic real data. Here is an example of the loss function in generative pre-training called Negative Log-Likelihood (NLL) loss~\cite{radford2018improving}:
\begin{equation}
    L_{gen} = -\sum_{i \in \mathcal{T}} \log P(x_i | x_{\setminus i}, \theta) \label{l_gen},
\end{equation}
where $\mathcal{T}$ is the set of target tokens to be predicted, $x_i$ represents the $i$-th token, $x_{\setminus i}$ represents all available context except token $x_i$, such as all tokens before the $x_i$ in autoregressive generation or visible tokens (unmasked positions) in reconstruction-based generation, $P(x_i | x_{\setminus i}, \theta)$ denotes the conditional probability assigned by the model to the correct token $x_i$, given the available tokens $x_{\setminus i}$ and model parameters $\theta$.
}

\subsec{Contrastive pre-training.} This method trains models to distinguish between data instances and learn robust data representations by evaluating similarities and differences among data samples~\cite{hadsell2006dimensionality}. A simple form of the contrastive loss for a pair of samples can be defined as follows:
\begin{equation}
    L_{con} = yD^2 + (1 - y)\max(0, \epsilon - D)^2, \label{l_con}
\end{equation}
where $y$ is a binary label indicating whether the pair has the same labels ($y=1$) or different labels ($y=0$), $D$ is the distance between the representations of the pair of samples, often calculated using a distance metric like Euclidean distance, and $\epsilon$ is a hyperparameter that defines the minimum distance between dissimilar pairs. Recent advancements~\cite{chen2020simple} extend contrastive learning to self-supervised paradigms, enabling the model to autonomously generate positive and negative labels from the data itself, thus eliminating the need for labeled datasets. This innovation significantly mitigates the dependency on manually labeled datasets, facilitating the processing of large-scale unlabeled data collections more efficiently.

\subsec{Hybrid pre-training.} This approach integrates multiple pre-training methods, including supervised, generative, and contrastive techniques~\cite{tian2023divide, kim2021hybrid}. 
It aims to tailor the training process to specific problem demands, leveraging the strengths of each method. 

Typically, supervised pre-training is used in scenarios with abundant labeled data.
However, since obtaining a large amount of high-quality labeled data can be expensive or impractical in many domains, generative and contrastive pre-training are frequently employed to fully leverage various types of datasets, improving the model's generalization ability and versatility. 

\subsubsection{Adaptation}
After pre-training, models can be customized to specific tasks or domains through \textbf{unimodal} or \textbf{cross-modal} adaptation depending on whether the data modalities are consistent between pre-training and adaptation stages.
The adaptation methods, primarily including model fine-tuning, prompt tuning, and prompt engineering, enhance the flexibility and power of UFMs across diverse application contexts. Selecting suitable adaptation strategies is crucial for maximizing the performance of pre-trained models on specific tasks and datasets while reducing resource costs.

\subsec{Model fine-tuning.} Model fine-tuning is a widely used adaptation method in UFMs. This approach involves additional training of the pre-trained model using a task-specific dataset~\cite{devlin2018bert, liu2019roberta}. The fine-tuning process adjusts the parameters of the model to enhance its effectiveness for the specific task. While fine-tuning is a straightforward and effective approach for task-specific adaptation, it often requires sufficient labeled data for the target task and substantial amounts of computational resources. Therefore, some parameter-efficient fine-tuning methods~\cite{hu2021lora} are presented to minimize the number of parameters that require adjustment during adaptation, maintaining a balance between model adaptation performance and resource consumption.

\subsec{Prompt tuning.} Diverging from the traditional ``Pre-train, Fine-tune'' paradigm, prompt-tuning employs lightweight prompt tokens containing relevant information about the target task. In practice, these prompts are often a small number of task-specific learnable parameters~\cite{jia2022visual}. During the prompt-tuning stage, the parameters of the pre-trained backbone model are frozen, and only the trainable prompts are adjusted. This method has proven effective in leveraging the inherent knowledge embedded in the pre-trained backbone models. Moreover, this ``Pre-train, Prompt and Predict'' paradigm~\cite{liu2023pre} is also more efficient, as it necessitates training only a minimal set of prompt parameters, thereby eliminating the need for extensive modifications to the model's existing parameters.

\subsec{Prompt engineering.} Prompt engineering~\cite{radford2019language} is a training-free approach to directly harness pre-trained foundation models like LLMs without altering their parameters. It involves crafting task-specific prompts that guide the model's responses. These prompts act as instructions or hints, steering the model's attention toward relevant information within the input data.

These adaptation methods, including model fine-tuning, prompt tuning, and prompt engineering, equip UFMs with the adaptability to excel in different urban contexts. The choice of adaptation method needs to comprehensively consider the data availability, computational resources, and task-specific requirements, and balance the adaptation performance and resource cost.

\section{Challenges of Building Urban Foundation Models} \label{challenge}
While UFMs offer significant potential for understanding and managing complex urban environments, building UFMs is a complex task that encounters several unique challenges. These challenges stem primarily from the inherent complexities of urban data, the dynamic nature of urban environments, the diverse urban task domains, and the concerns on privacy and security. Below, we delve into the key challenges that underscore the intricacies of building effective UFMs.

\subsec{Multi-source, multi-granularity, and multimodal data integration.}
The integration of multi-source, multi-granularity, and multimodal data is one of the primary challenges in building UFMs. UFMs necessitate effectively integrating data from various sources like disparate urban sensors, satellite imagery, social media feeds, and traffic systems. The collected data span varying spatial and temporal granularities, ranging from broad city-wide patterns to specific local details, and from annual records to real-time measurements. Moreover, the data modalities vary significantly, encompassing text, images, sensor readings, geovectors, \etc
Integrating these heterogeneous data types poses significant challenges in data preprocessing, normalization, and fusion, presenting obstacles for models to discern relevant patterns across diverse datasets and reconcile potential discrepancies or conflicts within the data.

\subsec{Spatio-temporal reasoning capability.}
Another major challenge in UFMs is mastering spatio-temporal reasoning. Urban environments are dynamic, with changes and patterns evolving over both space and time. UFMs need to understand and predict complex phenomena that are spatial-dependent and time-dependent. 
This involves sophisticated modeling of urban spatial distributions and temporal patterns, such as understanding geographical relationships between urban entities or activities across locations and the sequential, causal nature of urban temporal events or trends.
Advanced algorithms are required to handle
high-dimensional urban data and non-linear interactions, incorporating multiple spatial resolutions and temporal scales capturing the full spectrum of urban dynamics.
This spatio-temporal reasoning skill
is particularly crucial in urban contexts, however, it is often less emphasized or absent in established foundation models from other domains. 

\subsec{Versatility to diverse urban task domains.}
The versatility of UFMs to diverse urban task domains is another significant challenge. Urban environments are inherently complex, spanning a wide range of domains such as transportation, urban planning, energy management, and environmental monitoring. 
Each domain presents unique challenges and requirements, necessitating UFMs that are versatile enough to adapt to these varied contexts.
To meet this demand, UFMs are expected to seamlessly transition between tasks and domains, flexibly adjusting their analytical focus as needed without compromising on task performance. 
While such generalizability has emerged in LLMs, it remains a key limitation in existing UFMs due to the challenges, such as the heterogeneity of urban data and tasks, data sparsity, and distribution shifts across diverse spatial and temporal scenarios.


\subsec{Privacy and security concerns.}
The deployment of UFMs raises critical privacy and security challenges due to their reliance on extensive urban data from varied sources, like personal devices and social media. Ensuring compliance with privacy regulations, such as the General Data Protection Regulation (GDPR)~\cite{voigt2017eu}, is crucial, necessitating robust data anonymization, clear consent for data use, and transparent processing practices to protect individual privacy.
On the security front, UFMs must safeguard data integrity and prevent unauthorized access amidst diverse vulnerabilities inherent in multi-source data streams, \eg~protecting against spoofing attacks on sensor networks. 
Additionally, it should ensure model security to defend against various malicious activities, such as data privacy breaches, data poisoning, jailbreak attacks, adversarial attacks. Given their significant societal impact, effectively addressing these concerns is essential to developing UFMs that are both trustworthy and ethically responsible.

\begin{figure*}[tb]
\hspace{-3mm}
\includegraphics[width=1\columnwidth]{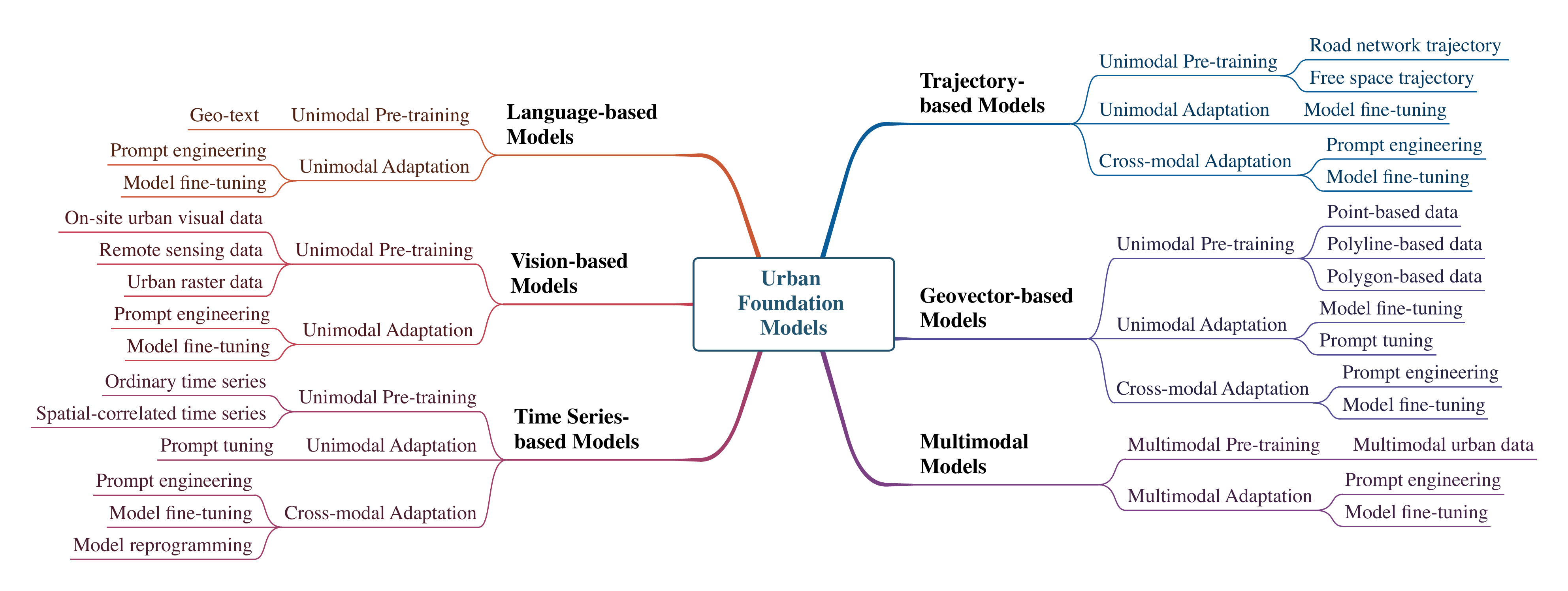}
\caption{A data-centric taxonomy for existing UFMs-related works based on the types of urban data modalities.}
\label{fig:taxonomy}
\end{figure*}

\section{Overview of Urban Foundation Models} \label{overview}
As illustrated in \figref{fig:taxonomy}, we introduce a data-centric taxonomy for the UFMs-related studies to shed light on the progress and efforts made in this field. Based on the urban data modalities that UFMs process, we categorize the existing works of UFMs into seven classes: language-based models,  vision-based models, time series-based models, trajectory-based models, geovector-based models, multimodal models, and others. After that, we introduce these studies through the lens of their focused pre-training and adaptation techniques. 

\subsection{Language-based Models}
Formally, text data can be represented as a sequence of tokens $ T = [t_1, t_2, \ldots, t_n] $ where each token $ t_i $ corresponds to a word or character in the text. These tokens are often transformed into numerical representations, like word embeddings, for processing. In recent years, extensive works have been done to develop Pre-trained Language Models (PLMs). PLMs typically leverage Transformer models as the backbone and are pre-trained over large-scale unlabeled text corpora, demonstrating strong capabilities in Natural Language Processing (NLP) tasks. Early representative PLMs, such as BERT~\cite{devlin2018bert}, XLNet~\cite{yang2019xlnet}, and GPT-series models~\cite{radford2018improving,radford2019language} (\ie GPT-1 and GPT-2), adopt a ``Pre-training and Fine-tuning'' paradigm for downstream tasks. Recently, researchers find that increasing the parameter scale of PLMs to a significant size (\eg tens of billions of parameters) not only enhances their capacity but also leads to remarkable emergent abilities, such as instruction following, in-context learning, and complex reasoning, which are not present in smaller PLMs~(\eg BERT). These enlarged PLMs are also termed LLMs in literature~\cite{zhao2023survey}. A notable progress of LLMs is the release of ChatGPT, which shows exceptional performance in human-like conversation and has garnered widespread attention from society.

People living in cities have generated a substantial corpus of textual data reflecting urban patterns, including documents (\eg traffic accident reports), dialogs (\eg ride-hailing conversations), and geo-texts (\eg geo-tagged tweets or entity descriptions). 
Unlike general text data, urban text is often tied to specific locations, with content reflecting geographic and spatial attributes, \eg landmarks, neighborhoods, and GPS coordinates. For example, a review of a restaurant on a social media platform is associated with the exact location of that restaurant. Additionally, urban text frequently exhibits a time-dependent nature, capturing real-time events and the evolution of urban dynamics (e.g., traffic incidents, public gatherings). These spatio-temporal characteristics make urban text a valuable source for studying the complex interactions among people, places, and events within a city.
While PLMs have shown proficiency in answering general questions across various domains, they struggle with location-specific and temporally correlated inquiries due to a lack of systematic injection of fine-grained urban knowledge.
Thus, it is crucial to adapt PLMs to better leverage and interpret the unique context of urban environments. Prior research in this area primarily falls into two categories: unimodal pre-training and unimodal adaptation approaches.

\subsubsection{Unimodal Pre-training}
Currently, extensive efforts have been directed toward developing general-purpose PLMs. These models are trained on text corpora that cover a wide range of topics and domains. While specialized urban datasets, such as traffic reports and social media posts, have been incorporated into the training process, the goal of them is to enhance the general capabilities of PLMs.
More recently, several studies~\cite{huang2022ernie,ding2023mgeo} focus on training PLMs from scratch by exclusively using urban textual data, achieving superior performance over generalist models on tasks within specific urban domains.
For instance, ERNIE-GeoL~\cite{huang2022ernie} is a pre-trained language model designed to improve various user services for Baidu Maps. 
ERNIE-GeoL is pre-trained on extensive data extracted from a heterogeneous graph encompassing rich urban and spatial knowledge, using masked language modeling and geocoding tasks. 
Similarly, Ding et al.~\cite{ding2023mgeo} propose MGeo, a language model specifically pre-trained on abundant geographic context data for query-POI matching. In specific, MGeo designs a geographical encoder to encode geographic context representations, and utilizes masked language modeling and contrastive learning in the pre-training stage.
However, large-scale pre-training often requires substantial computational resources and electricity, which may not be easy for researchers to obtain.

\subsubsection{Unimodal Adaptation}
Instead of training urban language models from scratch, previous studies have primarily focused on adapting established PLMs to urban scenarios.
In this way, we can make full use of the world knowledge embedded in established PLMs while significantly reducing the required computational cost.
\zhaobite{There are two categories of approaches: (1) prompt engineering and (2) model fine-tuning. }
Specifically, the adopted adaptation approaches are mainly two categories: prompt engineering and model fine-tuning. 

\subsec{Prompt engineering.}
Prompting aims to steer PLM's behavior toward desired outcomes via pre-defined text inputs, \eg~task description or a set of demonstrations, allowing PLMs to handle tasks they have never explicitly been trained for. Prompting offers a flexible, efficient, and user-friendly way to harness the capabilities of PLMs without the need for retraining or fine-tuning the model.
In recent literature, several preliminary studies suggest that PLMs have encoded a wealth of urban factual and spatio-temporal knowledge from their training corpora. These knowledge can be effectively accessed via prompt-engineering techniques. For instance, Gurnee~\cite{gurnee2023language} conducts empirical studies demonstrating that LLMs learn representations of space and time, which are resilient to changes of prompts and coherent across different urban geo-entities.
Likewise, Ji et al.~\cite{ji2023evaluating} indicate that LLMs are capable of representing geometric properties and relationships of spatial objects by using suitable prompts. In addition, Prabin et al.~\cite{bhandari2023large} investigate various geospatial skills of LLMs, including geospatial knowledge, awareness, and reasoning. By utilizing carefully curated prompts, LLMs can exploit the encoded geospatial knowledge to benefit diverse tasks, such as deriving coordinates of cities and understanding spatial prepositions. 

Moreover, several studies have applied LLMs to various urban perception and decision-making tasks, including urban concept comprehension~\cite{fu2023towards}, travel planning~\cite{roberts2023gpt4geo,xie2024travelplanner}, and geospatial question answering~\cite{mooney2023towards}. To name a few, Aghzal et al.~\cite{aghzal2023can} systematically evaluate the spatiotemporal reasoning abilities of LLMs in path planning tasks via different few-shot prompting strategies. GeoLLM~\cite{manvigeollm} focuses on extracting geospatial knowledge from LLMs by using prompts augmented with auxiliary map data, achieving superior performance in real-world geospatial prediction tasks. Zheng et al.~\cite{zheng2023chatgpt} explore the potential of LLMs in traffic safety applications through prompt information, including accident reports automation, traffic data augmentation, multisensory data analysis, and contrastive cross-modal learning. Mai et al.~\cite{mai2022towards,mai2023opportunities} perform a comprehensive study on the effectiveness of LLM prompting across a wide range of urban applications. With prompting, LLMs can act as zero-shot or few-shot learners, outperforming specialized models in tasks such as toponym recognition, country-level time series forecasting, and urban function classification. 

While direct prompting is flexible and resource-efficient, it demands powerful, generalist LLMs and rich contextual information to achieve satisfactory performance, which may restrict the applicability of LLMs in broad scenarios.

\subsec{Model fine-tuning.}
Generalist PLMs, pre-trained on vast amounts of web-based textual data, may lack sufficient knowledge for specific tasks or domains. Thus, some studies focus on fine-tuning PLMs with domain-specific data to enhance model performance in targeted areas. SpaBERT~\cite{li2022spabert} is a spatial language model designed for geo-entity comprehension. It extends BERT to capture the spatial varying semantics of geo-entities by fine-tuning on synthetic sentences from geographic databases. Additionally, SpaBERT introduces a spatial coordinate embedding module to preserve spatial relationships of geo-entities in latent feature space. GeoLM~\cite{li2023geolm} proposes to jointly learn linguistic and geospatial context via contrastive learning and masked language modeling. Compared with SpaBERT, GeoLM can effectively retain general linguistic information, which is crucial for many geospatial tasks.
Xie et al.~\cite{xie2023quert} introduce QUERT, a language model tailored for interpreting user search queries in the travel domain. Concretely, QUERT is jointly fine-tuned on several domain-specific tasks, such as geohash code prediction and geography-aware mask modeling, showcasing great performance improvements in various downstream travel-related tasks. 
More recently, Mei et al.~\cite{mei2023improving} introduce a fine-tuned BERT model to enhance query intent understanding in POI retrieval tasks. The model is initialized from BERT’s pre-trained weights and fine-tuned on both user behavior logs and the POI corpus. 

In addition, considerable efforts have been made to fine-tune open-source LLMs (\eg LLaMA~\cite{touvron2023llama}, ChatGLM~\cite{du2022glm}) for urban applications. 
In transportation scenarios, Wang et al.~\cite{wang2024transgpt} create the first transportation-focused LLM, named TransGPT-SM, by fine-tuning ChatGLM on textual data from diverse sources, including traffic documents, driving tests, and academic databases. TransGPT-SM demonstrates great potential in traffic-related tasks, such as traffic planning, question-answering, and traffic report generation.
Moreover, LLMLight~\cite{lai2023large} presents a pioneering framework that utilizes LLMs as control agents to enable human-like decision-making in traffic signal control tasks, showcasing remarkable effectiveness, interpretability, and generalization ability.
CoLLMLight~\cite{yuan2025collmlight} further extends this line of work by introducing cooperative LLM agents for network-wide traffic signal control, emphasizing inter-intersection coordination through structured spatio-temporal modeling and complexity-aware reasoning.
Zhou et al.~\cite{zhou2026llmnav} propose an LLM-powered cooperative framework for large-scale multi-vehicle navigation, highlighting coordinated decision-making among multiple agents in complex urban transportation environments.
Beyond transportation, Wang et al.~\cite{wang2023optimizing} develop an LLM in the field of urban renewal, which first generate domain-specific instruction data via the self-instruct~\cite{wang2022self} method, and then fine-tune ChatGLM on instruction data through Prefix~\cite{li2021prefix} and LoRA~\cite{hu2021lora} strategies.
PlanGPT~\cite{zhu2024plangpt} is proposed to improve the comprehension and reasoning capabilities of LLMs in urban and spatial planning domain. In particular, PlanGPT utilizes a two-stage fine-tuning approach, where the LLM is first trained on general text data and then fine-tuned on specialized urban planning datasets.
More recently, LAMP~\cite{balsebre2024lamp} presents a retrieval-enhanced fine-tuning framework to incorporate the knowledge of geo-entities within a specific city into LLMs. Through a POI retrieval task, LAMP creates a fine-tuning dataset encompassing the information of POIs and the spatial position of POIs and queries. 
CityGPT~\cite{feng2024citygpt} constructs a more comprehensive instruction-tuning dataset by simulating how humans navigate, interact, and reason in urban spaces. It demonstrates the effectiveness of this curated dataset across diverse urban scenarios by fine-tuning various LLMs.

Related efforts in geoscience and GIS also follow similar fine-tuning paradigms. Deng et al.~\cite{deng2024k2} introduce K2, a LLaMA-based model for geoscience knowledge understanding and utilization. K2 is fine-tuned on both unlabeled geoscience text corpus (5.5 billion tokens) and domain-specific supervised data. The model exhibits strong geoscience expertise, which significantly improves the performance of question answering and instruction following within the geoscience area. 
Along this direction, BB-GeoGPT~\cite{zhang2024bb} builds a GIS-oriented LLM through domain-specific pre-training and instruction tuning, enabling stronger understanding and utilization of geographic knowledge in downstream GIS tasks.

\rev{Looking ahead, advancing UFMs beyond basic language capabilities is essential for building truly intelligent urban systems. As discussed earlier, next-generation UFMs should integrate spatio-temporal reasoning, action-oriented planning, and tool-augmented interaction, enabling them to interpret not only static facts but also continuously evolving urban dynamics such as traffic conditions, environmental changes, and public events.
However, the integration of these advanced abilities also amplifies a fundamental challenge—hallucination~\cite{tonmoy2024comprehensive}. LLM-based UFMs may generate plausible but incorrect descriptions of urban states, infer nonexistent spatial relationships, or misinterpret time-varying patterns, which can cause substantial risks in safety-critical applications. Incorporating explicit grounding mechanisms, such as real-time data retrieval, sensor-aware tool usage, and verification modules, is therefore crucial to ensure factual consistency. In particular, robust tool utilization~\cite{qu2024toolsurvey} can mitigate hallucination by anchoring model outputs to authoritative data sources and operational systems, thereby enabling UFMs to interact with and respond to dynamic urban environments in a reliable and trustworthy manner.
}




\subsection{Vision-based Models}
Visual data encompasses any form of visual information captured in image or video formats, such as street view images, satellite imagery, and urban surveillance footage. Formally, an image can be represented as a 3-dimensional matrix, $ I(x, y, c) $ where $x$ and $y$ denote the spatial dimensions (width and height) of the image, and $c$ represents the color channels. Each entry in the matrix corresponds to a pixel value. 

Recent years have witnessed significant advances in the development of Vision Foundation Models (VFMs)~\cite{radford2021learning,kirillov2023segment,bai2023sequential}.
Early studies~\cite{chen2020simple,he2020momentum} focus on pre-training Convolution Neural Networks (CNNs) (\eg ResNet~\cite{he2016deep}) on ImageNet~\cite{deng2009imagenet} dataset for downstream tasks. Unfortunately, CNN-based architectures could struggle with scaling up to very large datasets due to their limited capacity~\cite{bai2023sequential}. Vision Transformer (ViT)~\cite{dosovitskiy2020image} was proposed to learn visual representations with Transformer architecture by representing an image as a sequence of image patches. Compared with CNN-based counterparts, ViT-based models have a much higher capacity to effectively absorb knowledge from vastly available data. 

Since the advent of ViT, numerous studies have utilized ViT to develop advanced VFMs. In existing literature, there are two categories of VFMs: language-augmented VFMs and vision-only VFMs. Language-augmented models, such as CLIP~\cite{radford2021learning}, aim to learn an image encoder by aligning paired text and images with contrastive learning. Conversely, vision-only VFMs are pre-trained on purely visual content data through self-supervised objectives, such as masked autoencoder (MAE)~\cite{he2022masked} and LVM~\cite{bai2023sequential}. These emerging VFMs, pre-trained on general visual data, exhibit robust generalization and impressive zero-shot learning abilities across a wide range of computer vision tasks. 
However, since the distributions of urban visual data (\eg satellite images) are significantly different from general visual data (\eg natural images)~\cite{jakubik2023foundation}, existing VFMs might not be suitable for handling urban-related tasks. Consequently, building tailored urban VFMs becomes highly necessary and practical. 
We will discuss representative techniques in the following sections.

\subsubsection{Unimodal Pre-training}
With the widespread deployment of camera and satellite technologies, an immense volume of visual data has been collected in urban spaces, capturing various aspects of city landscapes, infrastructures, public spaces, and human activities. The massive availability of data prompts a series of research to train large urban vision models from scratch. We categorize existing models based on the type of pre-training data they utilize: on-site urban visual data, remote sensing data, and urban raster data.

\subsec{On-site urban visual data.}
On-site urban visual data, such as street-view imagery~\cite{biljecki2021street} and surveillance footage~\cite{buch2011review}, are generated by ground-level devices (\eg smartphones, cameras, and automotive LiDARs) deployed in cities and closely tied to specific spatial locations, providing rich details of street scenes, vehicular and pedestrian traffic, \etc Such kind of data can be utilized to assess real property values~\cite{law2019take}, discover high-congestion areas~\cite{shi2023open}, and quantify the socioeconomic impact of cities~\cite{liu2023knowledge}. 

Early research in this field predominantly relies on task-specific supervision signals, such as demographic variables~\cite{gebru2017using}, to learn visual representations of urban imagery. However, these approaches incur additional labeling costs and often exhibit limited generalization capabilities. To address these issues, recent studies~\cite{wang2020urban2vec,li2022predicting,liu2023knowledge} have shifted towards self-supervised learning, which learns generalizable image representations from unlabeled visual data for a variety of downtream tasks. For instance, Urban2Vec~\cite{wang2020urban2vec} applies the Tobler’s First Law of Geography~\cite{miller2004tobler}—which states that “everything is related to everything else, but near things are more related than distant things” —to develop a contrastive learning framework. This framework learns representations of street-view images by enforcing that spatially adjacent images are similar in latent feature space. Another example, KnowCL~\cite{liu2023knowledge} employs contrastive loss to maximize the mutual information between a region's street-view image representation and its corresponding knowledge graph embedding. So far, existing studies mainly focus on relatively small datasets, and large-scale pre-training of VFMs on extensive on-site urban visual data remains underexplored.

\subsec{Remote sensing data.}
Remote Sensing (RS) data are typically captured from satellites, aircraft, or drones. RS can cover extensive urban areas in a single image or dataset, making it particularly useful for large-scale analysis, such as land use classification, urban planning, and environmental monitoring. Motivated by the success of self-supervised learning in computer vision, numerous studies~\cite{ayush2021geography,manas2021seasonal} have explored the application of this technique to extract knowledge from unlabeled RS data. More details about self-supervised learning algorithms for RS data can be found in~\cite{wang2022self}. 
Building on these works, recently the research on RS foundation models has been largely advanced, attracting widespread attention from both academia~\cite{dosovitskiy2020image,wang2022advancing} and industry~\cite{jakubik2023foundation}.

Most RS foundation models follow the Masked Image Modeling (MIM)~\cite{he2022masked} paradigm, which first encodes a masked RS image with an image encoder, and then decodes the unmasked portion to reconstruct the entire image. Unlike natural images, RS images usually contain rotated, dense, and small objects, and taking these unique properties into account during pre-training is beneficial.
Specifically, Wang et al.~\cite{wang2022advancing} train a plain ViT with 100 million parameters to solve various RS tasks, ranging from detecting urban objects to identifying region functions and understanding the whole city's landscape. Notably, a rotated, varied-size window attention is proposed to replace the standard attention block in ViT, significantly reducing the computation cost for high-resolution RS images. ScaleMAE~\cite{reed2023scale} introduces scale invariance into standard ViT and adopts a Laplacian-pyramid network to learn multi-scale semantics in RS images. More recently, Cha et al.~\cite{cha2023billion} investigate the impact of scaling up ViT to one billion parameters in the RS domain. Additionally, RingMo~\cite{sun2022ringmo} employs Swin Transformer~\cite{liu2021swin}, a well-known variant of ViT, as the model backbone for MIM-based pre-training. In particular, RingMo randomly preserves some pixels in masked patches, ensuring that critical fine-grained details are retained during image reconstruction. This strategy enables the model to effectively capture dense and small objects that are often overlooked in RS scenarios.
RingMo-Sense~\cite{yao2023ringmo} further extends RingMo for RS spatiotemporal prediction tasks. It develops a multi-branch structure to learn multi-scale spatiotemporal representations, and utilizes different masking strategies (\ie block-wise, tube-wise, and frame-wise masking) for distinct branch during MIM-based pre-training. Recently, several studies~\cite{mai2023csp,mendieta2023towards} start to explore different paradigms for training RS foundation models. To name a few, CSP~\cite{mai2023csp} learns visual representations by aligning paired locations and RS images through a contrastive pre-training objective. GFM~\cite{mendieta2023towards} constructs a RS foundation model using a multi-objective continual pre-training paradigm, which utilizes an ImageNet pre-trained model and distillation strategy to guide model training, while simultaneously learning domain-specific RS features through masked image modeling.

\subsec{Urban raster data.}
Urban raster data is organized in a grid-like structure, where each grid cell contains a value representing an attribute of interest. Such data often exhibit unique time series patterns, including trends and cyclic behaviors. Examples include air pollution patterns~\cite{zheng2013u}, meteorological data~\cite{chen2023foundation}, and regional mobility flow~\cite{zhang2017deep}. Urban raster data can be naturally represented as images, with pixel intensities characterizing the spatial distribution of specific variables, making it suitable for processing by vision models.

In recent years, the rapid development of VFMs has significantly impacted the processing paradigm of urban raster data. 
Large VFMs, pre-trained on massive, diverse urban raster datasets, demonstrate strong generalization capability and may be adapted for a broad range of analysis tasks, especially in climate and weather domains~\cite{pathak2022fourcastnet,bi2023accurate,nguyen2023climax,man2023w,chen2023fengwu,lam2023learning,ni2026uniextreme}. 
For instance, FourCastNet~\cite{pathak2022fourcastnet} pioneers developing a high-resolution global weather forecasting model based on ViT architecture and adaptive Fourier neural operator~\cite{guibas2021adaptive}. It is pre-trained on an extensive reanalysis of meteorological records using autoregressive objective.
Following this, Pangu-Weather~\cite{bi2023accurate} first represents surface and upper-air weather variables as 3D raster data, and divide 3D data into 3D patches. These patches are then passed through a 3D Earth-specific transformer for medium-range weather forecasting. Notably, Pangu-Weather outperforms the world’s leading Numerical Weather Prediction (NWP) system on reanalysis data, demonstrating significant potential for pre-training on raster-based weather data.
In contrast, GraphCast~\cite{lam2023learning} first organizes raster-based weather data into a multimesh graph, and then leverages graph neural networks to process the data with an encoder-decoder architecture.
Furthermore, FengWu~\cite{chen2023fengwu} treats weather modeling as a multi-modal multi-task problem, and develops a Transformer-based architecture that fuses various atmospheric variables/modalities to make independent predictions. Different from the above models trained in an autoregressive manner, W-MAE~\cite{man2023w} adapts MAE technique into weather and climate modeling. This model is pre-trained by reconstructing the spatial relations among meteorological variables, which allows it to be fine-tuned for enhanced performance in downstream forecasting tasks.
Beyond weather forecasting, Nguyen et al.~\cite{nguyen2023climax} present ClimaX, the first foundation model for broad climate and weather modeling tasks. ClimaX improves the standard ViT architecture with decoupled variable encoding and aggregation strategies, which enhance the model's adaptability to heterogeneous atmosphere variables. It also generalizes well to scenarios involving regions and variables not encountered during pre-training. More recently, Aurora~\cite{bodnar2024aurora} unifies the predictive modeling of air pollution, global ocean waves, tropical cyclone tracks, and weather by pre-training a Swin Tranformer~\cite{liu2021swin} on large-scale, heterogeneous Earth system data. The pre-trained Aurora model can be further fine-tuned for any desired Earth system prediction task, showcasing great potential as a foundation model for Earth system applications. Separately, UniExtreme~\cite{ni2026uniextreme} introduces a universal foundation model for forecasting on multiple types of extreme weathers, demonstrating robust prediction capability under high-impact and rare meteorological conditions.

\subsubsection{Unimodal Adaptation}
In data-constrained urban scenarios, we may not have sufficient domain-specific data to train a large-scale VFM from scratch. To address this issue, several studies directly adapt off-the-shelf VFMs, pre-trained on general visual data, for urban tasks. This section discusses how such adaptations can be effectively achieved through prompt engineering or model fine-tuning techniques.

\subsec{Prompt engineering.}
Current research in this category mainly focuses on adapting the Segment Anything Model (SAM)~\cite{kirillov2023segment} for urban RS image segmentation. SAM is a foundation model tailored for segmentation tasks, consisting of three components: (1) a ViT-based image encoder for generating image embeddings; (2) a prompt encoder that processes user prompts (specify what to segment in an image) to create embeddings; (3) a lightweight mask decoder that combines image and prompt embeddings to predict segmentation masks. Particularly, the user prompt can vary from sparse (\eg points, boxes, and texts) to dense (\eg coarse-grained masks). 
Since SAM is trained on natural images and struggles to generalize to RS scenes, we hereafter discuss specific prompted methods to effectively adapt SAM for RS datasets.
Wang et al.~\cite{wang2023samrs} propose SAMRS for RS segmentation by directly harnessing the zero-shot capabilities of SAM. SAMRS manually designs six basic prompts based on the characteristics of RS images, and identifies the optimal combination of prompts through experimentation for effective segmentation.
Build upon this, RSPrompter~\cite{chen2023rsprompter} further simplifies the process by automating prompt generation. It generates appropriate prompts, such as point or box embeddings, for SAM input by analyzing the hidden layers of the encoder.

Besides SAM-based approaches, recently Roberts et al.~\cite{roberts2023charting} explore the geographic and geospatial capabilities of multimodal LLMs (\eg GPT-4V) by using Chain of Thought (CoT) prompting~\cite{kojima2022large}. Through a series of experiments on various types of visual geographic data, the study demonstrates multimodal LLMs' impressive capabilities in extracting fine-grained details from urban imagery, but they struggle with tasks requiring precise localization and accurate drawing of bounding boxes.

\subsec{Model fine-tuning.}
In urban scenarios, model fine-tuning is used to refine a pre-trained VFM towards a specific capability, \eg~RS image segmentation.
For instance, GeoSAM~\cite{sultan2023geosam} enhances SAM's performance on mobility infrastructure segmentation through parameter-efficient fine-tuning, with the help of automatically generated visual prompts.
RingMo-SAM~\cite{yan2023ringmo} employs fine-tuning to update the prompt encoder of SAM for multi-source RS segmentation.
Moreover, recent studies~\cite{zhang2022migratable,haas2023learning} have explored the application of language-augmented VFMs (\eg CLIP) in street scene analysis.
Specifically, Zhang et al.~\cite{zhang2022migratable} introduce a method for street-view image analysis using a language-augmented VFM, fine-tuned with customized loss function and classification layers. This method facilitates the extraction of detailed textual descriptions from varied urban landscapes, linking street-view images with geo-tagged text. 
StreetCLIP~\cite{haas2023learning} adapts CLIP for street-view image geolocalization tasks by fine-tuning it with synthetic captions through a meta-learning approach.
In addition, GeoCLIP~\cite{vivanco2024geoclip} addresses the worldwide geo-localization by proposing a novel Image-to-GPS retrieval approach that aligns images with their corresponding GPS coordinate. To achieve this goal, GeoCLIP jointly fine-tunes a pre-trained CLIP-based image encoder with a newly added location encoder using location-image pairs.

Despite fruitful progress, existing studies primarily focus on task-specific applications using relatively small datasets, which restricts the models' versatile utility and adaptability across diverse urban contexts. To overcome this limitation, future research should prioritize cross-domain pre-training and cross-modal supervision. On the one hand, cross-domain pre-training can incorporate data from different urban visual sources, such as street-view and remote sensing images, to enhance the model's generalization capabilities. On the other hand, pre-training with cross-modal supervision, integrating data types such as POIs, time series, and trajectories, can enrich the models’ understanding of urban dynamics by capturing multi-modal semantic relationships within cities.



\subsection{Time Series-based Models} 
Time series is the collection of observable records in an temporally ordered manner, which represents the evolution of variable states over time. 
Time series data are ubiquitous in urban contexts due to the time-varying characteristic of urban environments, associated with various important urban tasks such as traffic forecasting~\cite{zhang2023irregular,zhang2020semi,han2024bigst}, air quality forecasting~\cite{han2022semi,han2023kill, han2023machine}, energy consumption prediction~\cite{xue2023promptcast}. 
Formally, a time series can be denoted as $\mathbf{X}\in\mathbb{R}^{T\times N\times C}$, where $T$ is the number of time steps, $N$ is the number of variables~(\eg~sensors), and $C$ represents the number of channels. 
In this section, we review the recent advances in building foundation models that can handle urban time series, classifying these methods into three categories: 1) unimodal pre-training, 2) unimodal adaptation, and 3) cross-modal adaptation.

\subsubsection{Unimodal Pre-training} Unimodal pre-training typically involves training deep neural networks on numerous large time series datasets in supervised or self-supervised manners. In pre-training, the model learns to capture the intricate temporal patterns and inter-variable correlations within the time series data. 
This section respectively illustrates the pre-training methods of foundation models on ordinary time series and spatial-correlated time series.

\subsec{Ordinary time series.} For ordinary time series, existing works focus on extracting the temporal patterns and dependencies during pre-training and can be categorized into: supervised, generative, contrastive, and hybrid methods, according to the training schemes.

Supervised approaches~\cite{oreshkin2021meta,rasul2023lag,dooley2024forecastpfn,liu2024generative,liu2024timer,ekambaram2024ttms,dasdecoder} directly pre-train the model with ground truth information (\eg future values for forecasting, labels for classification). 
To name a few, GPHT~\cite{liu2024generative} delves into the auto-regressive pre-training strategy similar to language modeling, which could sufficiently capture the temporal patterns within the predicted series and generalize to arbitrary horizon lengths.
Recently, TimesFM~\cite{dasdecoder} explores building a time series foundation model with good zero-shot forecasting performances, which constructs a large time series corpus and designs a decoder-only model architecture.
Time-MoE~\cite{shi2024timemoe} resorts to mixture-of-expert (MoE)~\cite{shazeer2017outrageously} techniques to scale time series foundation models up to billion-level parameters that achieves a trade-off between model capacity and inference overhead.

Generative approaches usually use a masked reconstruction scheme that adds random masking to input time series and trains the model to recover the missing contents~\cite{zerveas2021transformer}. 
For instance, SimMTM~\cite{dong2024simmtm} finds directly masking a portion of time series will seriously ruin the temporal semantic variations. It regards masked reconstruction from a manifold perspective and presents to recover the original time series from multiple masked series. 
Besides, many other masked reconstruction-based pre-training approaches~\cite{nie2023time,ekambaram2023tsmixer,liu2023pt} focus on improving the model architecture for better time series analysis. 
Lately, based on the decoupled auto-encoder architecture~\cite{cheng2023timemae}, HiMTM~\cite{zhao2024himtm} has devoted the pioneering efforts to integrating multi-scale temporal pattern extraction into masked time series reconstruction. 
Inspired by contrastive learning, TimSiam~\cite{dong2024timesiam} designs a novel reconstruction-based pre-training strategy, which models the correlation among distanced subseries beyond the current context such that capture the global dynamics of the whole time series. 
Moreover, Moirai~\cite{woo2024moirai} builds a masked-encoder-based versatile Transformer architecture to accommodate time series with varied frequencies, distinct variate sizes, and diverging distributions.

Contrastive approaches aim to enhance the consistency between the representations of time series objects that share similar semantic meaning (\eg~positive pairs) and reduce that for those with disparate semantics (\eg~negative pairs).
The key to contrastive-based time series pre-training lies in defining the consistency of positive pairs that should be drawing close to each other. There are primary five types of consistency adopted for contrastive time series pre-training~\cite{ma2023survey}: subseries consistency~\cite{franceschi2019unsupervised} assumes the representations of a reference time series should be similar to its subseries; transformation consistency~\cite{eldele2021time,woo2022cost} indicates time series of distinct augmentation views should be consistent; temporal consistency~\cite{tonekaboni2021unsupervised} means that adjacent subseries in a time series should be similar; contextual consistency~\cite{yue2022ts2vec} assumes that the representations of the same timestamp in two augmented view of the same subseries should be consistent; consecutive consistency~\cite{ragab2022self,eldele2021time,zheng2023simts} implies that the two consecutive parts, history and horizon, of a time series should be consistent. 

Hybrid approaches~\cite{liu2024unitime,gao2024units} integrate the aforementioned training schemes for robust pre-training. For example, 
to capture the variations of temporal characteristics over source domains~(\eg weather, electricity), UniTime~\cite{liu2024unitime} empowers the input embedding with domain-specific text information as the manual prompt. Besides, it utilizes input and output padding to adapt varied data characteristics (\eg length of history and horizon), and designs input masking mechanism for imbalanced learning that is robust to varying convergence rates. 
UniTS~\cite{gao2024units} modifies the Transformer to accommodate the variability of temporal dynamics among multi-domain time series data. It supports multi-task supervised training or unified masked reconstruction pre-training, and can be adapted to diverse time series tasks via prompt learning.

\subsec{Spatial-correlated time series.} 
Beyond ordinary time series pre-training that usually only considers temporal dimension, the studies on urban time series emphasize incorporating the spatial dimension to explicitly accommodate the dependencies between variables. Much relevant literature also name it as spatio-temporal data~\cite{yuan2024unist,liu2022contrastive,shao2022pre}.

Some works~\cite{wang2018cross,yao2019learning,zhao2022st} model the spatial relations as a grid map with fixed sizes and employ CNNs to encode the spatial pattern. For example, ST-GSP~\cite{zhao2022st} proposes a semantic flow encoder with ResNet as the backbone to capture the spatial dependencies. Besides, it devises a Transformer encoder with an early-fusion paradigm to capture multi-scale temporal information for reconstruction-based pre-training and downstream urban flow prediction. Lately, UniST~\cite{yuan2024unist} further proposes a universal foundation model for spatio-temporal prediction that could generalize diverse data formats and variations of data distributions. It designs a generative pre-training strategy with four distinct masking techniques and a prior knowledge-guided prompt network for adaptation.
However, grid-based spatial modeling is limited to local regular dependencies and fails to exploit irregular and arbitrary variable-wise connections. 

In light of that, recent works~\cite{pan2019urban,jin2022selective,lu2022spatio,tang2022domain,liu2022contrastive} focus on graph-based spatial encoding and usually use Spatio-Temporal Graph Neural Networks (STGNNs) to extract the spatio-temporal information. For instance, due to data scarcity, STGCL~\cite{liu2022contrastive} devises a contrastive pre-training framework for traffic forecasting. It introduces node- and graph-level contrast paradigms, four data augmentation techniques, and debiased negative filtering strategies from both temporal and spatial views. 
Opencity~\cite{li2024opencity} crafts an adaptive foundation model for traffic prediction by handling the spatio-temporal heterogeneity from diverse data characteristics.
However, these methods adopt pre-defined graph structures that may be biased and noisy, which negatively impact spatial information extraction~\cite{shao2022pre} or cross-city knowledge transfer~\cite{jin2023transferable}.

To address the aforementioned drawbacks, STEP~\cite{shao2022pre} introduces a graph structure learning module in downstream forecasting. 
As vanilla STGNNs typically focus on short-term time series, STEP devises TSFormer for time series pre-training to capture long-term temporal patterns with segment-level representations.
TransGTR~\cite{jin2023transferable} focuses on learning the graph structure in a transferable way, which pre-trains a structure generator and transfers both the generator and the spatiotemporal forecasting module across cities. 
Further, MC-STL~\cite{zhang2023mask} spots the spatiotemporal heterogeneity (i.e., spatial correlation and region importance are time-varied), and proposes two pre-training strategies: spatial masked reconstruction and temporal contrastive learning, to understand inter-region correlations and regional pattern variations over time. 
Similar to TransGTR, TPB~\cite{liu2023cross} focuses on cross-city transfer learning for spatio-temporal forecasting. To capture the traffic patterns in data-rich cities for transferring to data-scarce cities, TPB generates the representative traffic patterns via clustering, and conducts pattern aggregation during downstream forecasting. Additionally, it devises an attention-based graph reconstruction followed by STGNN for spatio-temporal pattern extraction. 
GPT-ST~\cite{li2023gpt} observes that existing methods encounter the lack of customized representations of spatio-temporal patterns and insufficient consideration of different levels of spatial dependencies. Thus, it pre-trains a spatio-temporal mask autoencoder by incorporating a parameter customization scheme and adaptive mask strategy, utilizing a hierarchical hypergraph structure to capture multi-level spatial dependencies from a global perspective.
\rev{In particular, to improve the computational and data efficiency, CompactST~\cite{han2025scalable} is developed as a compact spatio-temporal foundation model~($\sim$300K parameters) that can be efficiently pre-trained on massive, multi-domain urban datasets using a multi-scale spatio-temporal mixer and mixture-of-normalizers to handle heterogeneous patterns. The results show better prediction accuracy and efficiency, especially when downstream data is scarce.}



\subsubsection{Unimodal Adaptation} 
Unimodal adaptation aims to tailor the pre-trained time series foundation models to new time series datasets to enhance downstream task performance. 
Due to the emergence of prompt tuning methods in data modalities like natural language, vision, and graph~\cite{sun2023all}, prompting methods for time series also receive growing attention. For example, PT-Tuning~\cite{liu2023pt} adds tunable prompts into masked tokens to achieve the consistency of task objectives between pre-training (\ie masked reconstruction) and adaptation (\ie forecasting) in urban time series data. For time series classification, POND~\cite{wang2023prompt} further designs an instance-level prompt generator for adaptive prompting, besides the common prompt. Also, to capture diverse domain-specific information, POND proposes a meta-learning paradigm as well as elaborated contrastive constraints for prompt initialization.

Moreover, there have been some studies on prompt tuning for spatial-correlated time series tasks. To name a few, PromptST~\cite{zhang2023promptst} adds tunable prompts to spatiotemporal data to efficiently fine-tune the pre-trained traffic forecasting model and avoid catastrophic forgetting issue. STGP~\cite{hu2024prompt} designs a tow-stage prompting pipeline for spatio-temporal graph learning, which enables the prompts to curate domain knowledge and task-specific properties.
UniST~\cite{yuan2024unist} presents to learn spatial and a temporal memory pools with key-value structured parameters. Then the knowledge-guided prompts are extracted by querying the prompt memory pools with input representations.
MetePFL~\cite{chen2023prompt} and FedWing~\cite{chen2023spatial} focus on federated weather forecasting and use prompting techniques for personalized federated learning to reduce communication overheads and enhance privacy protection.


\subsubsection{Cross-modal Adaptation} 
Due to insufficient time series data to train a foundation model from scratch, cross-modal adaptation approaches directly adopt the foundation models pre-trained on other modalities and aim to transfer the modality-sharing knowledge to time series tasks.
Recently, most cross-modal studies have focused on adapting pre-trained LLMs for time series to sufficiently leverage their remarkable abilities~\cite{jin2023large,jin2024position}. 
Upon whether the LLMs are purely black-box or open-sourced, we categorize them into two types: embedding-invisible LLM (\eg ChatGPT, GPT-4\footnote{\url{https://openai.com/blog/gpt-4-api-general-availability}}) and embedding-visible LLM (\eg GPT-2~\cite{radford2019language}, LLaMA~\cite{touvron2023llama}), which follows~\cite{jin2023large}. For the former type of LLMs, we could just infer by prompting with API calls, while for the latter type, we can fine-tune the model parameters or reprogram the time series inputs in various tasks.

\subsec{Prompt engineering.} Some recently released LLMs hold striking capabilities but are closed-sourced (\ie embedding-invisible). A plausible solution for adaptation is to transform the time series into natural language sentences as the prompts for API calls. PromptCast~\cite{xue2023promptcast} first presents a prompt-based forecasting paradigm and releases a large-scale dataset PISA constructed with template-based prompts including weather temperature, electricity consumption and human mobility prediction tasks. Further, LLMTime~\cite{gruver2023large} contends that LLMs can be used directly as zero-shot forecasters without any added text or prompt engineering if the input time series data are carefully tokenized.

\subsec{Model fine-tuning.} Beyond prompting LLMs, many works have contributed to fine-tuning embedding-visible LLMs for time series tasks. FPT~\cite{zhou2023one1_nips} adopts pre-trained LLMs for diverse time series tasks. It only fine-tunes a small part of the parameters of LLM (\ie positional embedding and layer normalization layer) for efficiency, and uses linear probing and patching to align the time series of various urban domains (\eg~weather, electricity, traffic) to language modality. 
To align the modalities of time series and language, LLM4TS~\cite{chang2023llm4ts} design a two-stage fine-tuning process: firstly adopts auto-regressive supervised fine-tuning for time series; then fine-tunes the pre-trained model for downstream time series forecasting. Further, TEMPO~\cite{cao2023tempo} addresses LLMs' limitations in capturing time series features like trends, seasonality, and distribution shifts. It employs trend-seasonality decomposition for input time series, feeding these components into the LLM for reasoning, while tunable prompts mitigate distribution shifts across diverse datasets. 
To tackle time series with diverse context lengths, AutoTimes~\cite{liu2024autotimes} proposes an auto-regressive LLM-based forecasting approach, and adopts next token prediction as the training objective which is consistent with the LLM acquisition. Besides, AutoTimes designs token-wise prompting technique that utilizes the LLM generate the prompts indicating timestamp information.

To adapt the LLMs for spatial-correlated time series, GATGPT~\cite{chen2023gatgpt} uses a graph attention module for spatial graph encoding and passes the spatio-temporal embeddings to the pre-trained GPT-2 model for downstream spatio-temporal imputation. TPLLM~\cite{ren2024tpllm} adopts 1-D convolutional neural networks and graph convolutional networks to enable LLMs to understand spatio-temporal patterns in traffic data.
Additionally, STG-LLM~\cite{liu2024can} proposes a spatio-temporal tokenizer to transform spatio-temporal data into tokens and an adapter architecture for efficient adaptation, which facilitate the understanding of pre-trained LLMs. Similarly, ST-LLM~\cite{liu2024spatial} treats the time steps at each spatial location as tokens, and fuse various spatial and temporal prior information into input embedding to facilitate traffic prediction. 
\rev{Moreover, STD-PLM~\cite{huang2025std} fine-tunes a pretrained language model by partially updating attention, position embeddings, and layer-norm layers—using LoRA for efficient adaptation—to encode spatial-temporal structures. This selective tuning enables the pretrained language model to understand node topology, temporal patterns, and missingness, supporting forecasting, imputation, and strong zero- and few-shot generalization.}

\subsec{Model reprogramming.} To explicitly align time series to other modalities like natural language or voice,  model reprogramming~\cite{chen2024model} is a prospective way that neither edits the time series inputs directly nor fine-tunes the pre-trained models. 
For instance, V2S~\cite{yang2021voice2series} leverages a pre-trained acoustic model for time series classification through input reprogramming and output label mapping. 
TEST~\cite{sun2023test} presents a data-centric framework that freezes LLM parameters and refines inputs. It aligns time series embeddings to text representation using contrastive learning and employs tunable prompts for downstream forecasting and classification.
To unleash the potential of LLMs for time series forecasting, Time-LLM~\cite{jin2024time} reprogram time series into text prototype representations and add language-based prompts that describe the statistics of time series data. To avoid the large and dense reprogramming space, Time-LLM only maintains a small collection of the vocabulary by linearly probing the pretrained word embeddings.
To enhance the reprogramming framework for spatio-temporal data, RePST~\cite{wang2024empowering} decouples the spatio-temporal data in the frequency space to capture intricate spatio-temporal dependencies. Also, RePST discretely samples word embeddings from the whole vocabulary to avoid semantic mixing and enhance the expressivity of text prototypes for spatio-temporal data.
To align the spatio-temporal dependencies with the comprehension space of LLMs, UrbanGPT~\cite{li2024urbangpt} presents a spatio-temporal instruction tuning paradigm to align spatio-temporal data with the knowledge space of LLMs.

Despite the endeavors on both unimodal and cross-modal approaches, most existing time series-based UFMs are limited to homogeneous data formats (\eg~regularly sampled time series) and have insufficient zero-shot adaptation ability on new domains and tasks.
In addition, how LLMs effectively understand urban time series has never been sufficiently explored, which potentially facilitates more enhanced pre-training or adaptation techniques. 
Furthermore, incorporating other modal data, \eg~contextual text, into time series-based UFMs to enhance their versatile time series analysis abilities, such as time series question answers, is still under-explored.

\subsection{Trajectory-based Models} 
Trajectory data is an important modality prevalently existing in location-based urban tasks, such as driving behavior analysis~\cite{stpt2023, pim2021, cacsr2023, lai2023preference} and human mobility analysis~\cite{mmtec2023, hmtrl2022, start2023, movesim2020, ctle2021}. Formally, trajectory data is a specific type of geo-sensory data that can be defined as a sequence of spatio-temporal points, $ T = \left\{ (p_1, t_1), (p_2, t_2), \ldots, (p_n, t_n) \right\} $, where each point $ p_i $ is a location in space (often represented as geographic coordinates) and $ t_i $ is the corresponding timestamp. 
Trajectory-based UFMs are designed to capture the inherent spatio-temporal correlations and general mobility patterns among trajectory data via pre-training and adaptation techniques, enabling their applications to a broad range of urban contexts. 


\subsubsection{Unimodal Pre-training}

Trajectory data typically requires tailored modeling to suit specific applications~\cite{lai2023preference, movesim2020,zhu2023difftraj}. Unimodal pre-training, usually involving self-supervised learning on large pure trajectory datasets, enables models to capture intrinsic features and patterns among trajectories, enhancing their generalizability across different application scenarios. We introduce the pre-training methods on two types of trajectory data: road network trajectories and free space trajectories.

\subsec{Road network trajectory.}
Road network trajectory refers to trajectories that are confined within road networks, \eg~the paths of vehicles or pedestrians on streets and highways. Learning low-dimensional embeddings of trajectories and paths is one of the most crucial tasks in road network trajectory pre-training, as these embeddings can benefit various downstream tasks, such as travel time estimation and route planning. 
T2vec~\cite{t2vec2018} and Traj2vec~\cite{traj2vec2017} represent early attempts at trajectory-based pre-training models, adopting the sequence-to-sequence~\cite{seq2seq} encoder-decoder framework to learn trajectory representations. These models are pre-trained by reconstructing trajectory sequences through a generative self-supervised task.
However, these methods process raw trajectories as input, potentially compromising the quality of trajectory representation due to noise. To mitigate this, Trembr~\cite{Trembr2020} projects raw trajectories onto road networks, using road segment sequences as inputs. Furthermore, the decoder in Trembr is tailored to reconstruct both road segments and their associated travel times. This enhancement enables the effective use of learned trajectory representations in various downstream applications, such as trajectory similarity measurement, travel time prediction, and destination forecasting.

Recently years, more advanced methods are presented to further refine trajectory representations. 
PIM~\cite{pim2021} is a two-stage trajectory representation model employing node2vec for generating road embeddings. It utilizes mutual information maximization, a contrastive learning-based method, to train the LSTM encoder for route embeddings.
Unlike PIM, LightPath~\cite{lightpath2023} directly learns path representations for diverse downstream tasks by applying a self-supervised relational reasoning method to expedite the training of path encoders.
STPT~\cite{stpt2023} presents a spatial-temporal pre-training and fine-tuning approach for modeling human trajectory representation. 
In pre-training, it predicts the source of sub-trajectories through a self-supervised similarity learning task. The fine-tuning phase adapts the pre-trained model to diverse tasks using task-specific adapters. 
JGRM~\cite{ma2024more} proposes an approach that enhances trajectory representation by jointly modeling GPS and route trajectories. It leverages bi-modal information interaction, combining the rich details of GPS trajectories with the robust state transition records of route trajectories, and is pre-trained using self-supervised tasks, including mask language modeling and cross-modal matching. 

While the aforementioned methods utilize generative or contrastive approaches in pre-training, MMTEC~\cite{mmtec2023} contends that these methods could introduce biases in trajectory embeddings incurred by the pretext tasks. To mitigate this, MMTEC employs a unique pre-training task using maximum entropy coding to reduce biases in trajectory embeddings. 
This strategy facilitates the development of versatile, high-quality trajectory embeddings for various downstream tasks, such as similar trajectory search, travel time estimation, and destination prediction.
Additionally, some studies explore pre-training methods that combine both generative and contrastive approaches.
HMTRL~\cite{hmtrl2022} is a route representation learning framework tailored for multi-modal transportation recommendations. It integrates a spatiotemporal graph neural network to capture spatial and temporal autocorrelations and an attentive module for semantic coherence in route sequences. It also employs a hierarchical multi-task learning module for route representations across transport modes. The framework's spatiotemporal pre-training strategy, featuring masked attribute prediction and trajectory contrastive learning, bolsters its generalization capabilities.
START~\cite{start2023} introduces two self-supervised tasks for pre-training a trajectory representation model: span-masked trajectory recovery and trajectory contrastive learning, showing robust transferability across varied road network datasets and excelling in three distinct downstream tasks.
UniTraj~\cite{zhu2024unitraj} presents a universal trajectory foundation model designed to tackle task specificity, regional dependency, and data quality sensitivity in human trajectory modeling. It is pre-trained on WorldTrace, a global dataset with 2.45 million trajectories from 70 countries, and uses a general encoder-decoder structure with resampling and masking strategies during pre-training. The model demonstrates superior scalability and adaptability across various trajectory-related tasks such as trajectory recovery, prediction, classification, and generation.

\subsec{Free space trajectory.} 
Free space trajectory refers to trajectories in open, unstructured spaces without the constraints of road networks, such as pedestrian movements in parks or open areas. A representative subset of this category is check-in trajectories, where individuals record their location at specific POIs, often using social media or location-based services. 
To address the challenges in free space trajectory analysis, several innovative approaches have been developed, focusing on simulating and understanding human mobility in unstructured environments. Feng et al.~\cite{movesim2020} introduce MoveSim, a generative adversarial framework designed for simulating human mobility. This framework utilizes unique pre-training tasks for both the generator and discriminator, capitalizing on spatial continuity and temporal periodicity in human mobility, thereby markedly improving simulation fidelity.
SML~\cite{sml2021} is a framework developed for modeling human mobility. By utilizing spatiotemporal context learning and integrating common data augmentation techniques, it addresses data sparsity, thereby improving its capability to understand human mobility patterns accurately. Additionally, the authors implement a pre-training task for model pre-training, followed by fine-tuning through contrastive trajectory learning. 
CTLE~\cite{ctle2021} presents a pre-training model tailored to learn representations of locations, primarily for tasks such as next location prediction. This model uses the masked language model pre-training task from BERT, and generates location embeddings by considering contextual neighbors within trajectories, effectively addressing the multi-functional properties of locations.
Gong et al.~\cite{cacsr2023} propose CACSR, an adversarial contrastive model for learning check-in sequence representations. The model features a contrastive learning-based pre-training task for its encoder, thereby improving representation learning. After pre-training, CACSR is fine-tuned for diverse multi-classification problems, including location prediction and trajectory-user link tasks, showcasing notable performance.


\subsubsection{Unimodal Adaptation} 
Unimodal adaptation primarily focuses on fine-tuning the pre-trained trajectory models across different trajectory datasets, \eg~adapting a model trained on one city to another.
Reformd~\cite{Reformd2021} employs standard transfer learning procedures, transferring model parameters learned on a data-rich region to a target region with scarce data. This method models the invariant region dynamics in mobility prediction tasks within the origin region via a marked temporal point process. It concurrently learns the time interval and spatial distance distributions of check-in sequences as two distinct independent normalizing flows. During the transfer process, the spatial and temporal flows that have been learned are transferred, followed by fine-tuning the model parameters for the target region, incorporating an attention parameter for enhanced specificity.
Axolotl~\cite{Axolotl2022} also focuses on cross-region model transfer to enhance POI recommendations in data-scarce regions by leveraging data from data-rich regions. The method employs cluster-based knowledge transfer without the need for any common users or POIs between regions.
Building on this idea, CATUS~\cite{CATUS2023} addresses data scarcity and imbalance in POI recommendations across cities by transferring category-level universal transition knowledge. It achieves this through two self-supervised tasks in the pre-training phase: next category prediction and next POI prediction. During fine-tuning, CATUS uses a distance-oriented sampler better to align POI representations with a city's local context, taking into account the geographical distance and semantic similarity between POIs.
\rev{TransferTraj~\cite{wei2025transfertraj} introduces a unified trajectory learning model achieving both region and task transferability. Through a region-transferable encoder and a masking-based input–output scheme, it supports multiple trajectory tasks without retraining and outperforms existing baselines across zero-shot and few-shot settings.
Additionally, some studies explore adaptation strategies to improve generalization. For example, AdaMove~\cite{han2025adamove} proposes a preference-aware test-time adaptation mechanism that updates a trained model's parameters using the incoming test trajectory, enabling the model to better accommodate distribution shifts during inference.}

\subsubsection{Cross-modal Adaptation} 
Cross-modal adaptation approaches aim to adapt learned patterns, features, or knowledge from other data modalities to trajectory. Recent advancements in LLMs (\eg GPT-4~\cite{openai2023gpt4}, Llama2~\cite{touvron2023llama2}) have demonstrated exceptional zero-shot learning and generalization capabilities across various domains. A growing number of studies have explored to harness LLMs for trajectory-based tasks.

\subsec{Prompt engineering.} 
Most of these works leverage LLMs to tackle human mobility-related tasks through prompt engineering techniques. 
For instance, LLM-Mob~\cite{llmmob2023} designs effective prompts to enhance LLMs' reasoning on mobility data, producing both accurate and interpretable results on human mobility prediction tasks. To capture long-term and short-term dependencies, the mobility data is formatted into historical stays and context stays. 
Zhang et al.~\cite{zhang2023large} conduct empirical research to assess LLMs' effectiveness and limitations in identifying anomalous behaviors in human mobility data, demonstrating their performance compared with specialized anomaly detection algorithms.
LLM-MPE~\cite{llmmpe2023} seeks to use LLMs for forecasting human mobility in complex scenarios, \eg~public events, employing a chain-of-thought prompting approach to base predictions on historical mobility patterns and event descriptions.
\rev{Instead of directly prompting LLMs, TrajCogn~\cite{zhou2024trajcogn} designs a trajectory prompt that encodes spatio-temporal features, movement patterns, and travel purposes, and introduces a trajectory semantic embedder to map continuous trajectory signals into a language-compatible representation. It enables LLMs to perform multiple trajectory-related tasks, such as travel time estimation, destination prediction, and trajectory similarity search.}

Additionally, within the realm of autonomous driving, initiatives have been undertaken to harness LLMs for the purposes of motion planning and trajectory forecasting.
Keysan et al.~\cite{keysan2023can} present an approach that integrates descriptions of scenes in text form with pre-trained language encoders, aimed at predicting trajectories in the context of autonomous driving.
Similarly, GPT-Driver~\cite{gptdriver2023} redefines motion planning as a language modeling task, converting heterogeneous planner inputs into language tokens. This process employs a ChatGPT model to process the input tokens and generate driving trajectory predictions.
LanguageMPC~\cite{languagempc2023} employs LLMs such as ChatGPT to direct model predictive control (MPC) in autonomous driving. It uses prompts to feed environmental information to the LLM and leverages chain-of-thought in the dialogue to decompose the problem into manageable sub-problems.
DrPlanner~\cite{lin2024drplanner} introduces a framework that utilizes LLMs to refine motion planners, aiming to reduce the required human intervention. The framework employs LLMs to identify issues autonomously and suggest repairs for motion planners through structured descriptions, iteratively refining the process based on feedback from evaluations.

\subsec{Model fine-tuning and reprogramming.} 
Beyond prompt engineering, a few research explore fine-tuning or reprogramming the LLMs to fully harness them to trajectory-based tasks. For example, AuxMobLCast~\cite{auxmoblcast2022} converts numerical temporal sequences into natural language sentences and fine-tunes pre-trained language models such as BERT, RoBERTa, GPT-2, and XLNet to predict visitor counts at POIs. This work provides empirical evidence of pre-trained language models in discovering sequential patterns for mobility forecasting tasks effectively.
Traj-LLM~\cite{lan2024traj} introduces a trajectory prediction framework using pre-trained LLMs without explicit prompt engineering. It employs sparse context encoding, parameter-efficient fine-tuning techniques for scene interaction learning, and a lane-aware probabilistic learning mechanism with the Mamba module for enhanced scene comprehension. In addition, a multi-modal Laplace decoder ensures diverse, scene-compliant predictions, achieving high accuracy and adaptability, even with limited data. 
\rev{
PLMTrajRec~\cite{wei2024plmtrajrec} fine-tunes a pre-trained language model using only limited dense trajectories and handles varying sampling intervals by converting interval and movement features into natural-language prompts. It unifies trajectories to a shared temporal scale, models road conditions through area-level traffic-flow prompts and demonstrates scalability and generalization abilities on trajectory recovery.

Besides model fine-tuning, Mobility-LLM~\cite{gong2024mobility} introduces a unified framework that reprograms LLMs to understand human mobility data by converting check-in sequences into semantically meaningful inputs. By designing several key modules for enriching POI semantics, modeling short-term visiting intentions, and capturing long-term travel preferences, it allows the LLM to interpret mobility behavior as if it were natural language.}

\rev{While exiting trajectory-based UFMs show strong potential for various trajectory-related applications, they primarily focus on pre-trained models with relatively small parameter sizes, and there is limited exploration into whether scaling up model and data size, as seen in LLMs, could unlock new capabilities for handling complex trajectory data. 
Moreover, current research primarily focuses on harnessing LLMs with direct prompting or simple fine-tuning. However, further exploration of more advanced fine-tuning strategies, such as instruction tuning and reinforcement learning fine-tuning, is expected to more effectively adapt LLMs to diverse trajectory-related tasks. These gaps highlight significant opportunities for the advancement of trajectory-based UFMs.}

\subsection{Geovector-based Models} 
Geovector data encompasses information that identifies the geographic locations of features and boundaries on urban maps, typically represented using coordinate systems~\cite{tempelmeier2021geovectors,zhang2024veccity,chen2024self,mai2023opportunities}. 
It is utilized to analyze spatial patterns and relationships across various urban tasks, such as location embedding~\cite{li2022spabert, gao2022geobert, wan2021pre, tempelmeier2021geovectors}, road network representation~\cite{zhang2023road,ma2024more}, and urban planning~\cite{zhou2023heterogeneous, li2023urban}. 
Formally, geovector data consists of three primary geometric elements: points (\eg POIs), polylines (\eg~road segments), and polygons (\eg administrative regions).
In geovector data, a point is represented as $p = (l, x)$, where $l$ denotes the geographical coordinates (latitude and longitude), and $x$ refers to its associated features (\eg~attributes, readings).
A polyline comprises a sequence of connected line segments, each defined by location coordinates. 
It is represented by a list of points $L = [(l_1, x_1), \cdots, (l_n, x_n)]$, where $l_i$ denotes the geographical coordinates of the $i$-th point, and $x_i$ denotes its associated features.
A polygon represents a land parcel, defined by a closed boundary formed by a sequence of connected polylines. It is denoted by $G = [L_1, L_2, \cdots, L_n]$, where $L_i$ represents the $i$-th polyline, enclosing a specific area.
This section reviews literature on building geovector-based UFMs, grouping the methods into three categories: (1) unimodal pre-training, (2) unimodal adaptation, and (3) cross-modal adaptation.

\subsubsection{Unimodal Pre-training}
Unimodal pre-training usually involves learning representations of geovector data entities (\eg POIs, road segments, or areas) through supervised or self-supervised tasks. This process captures key attributes and relationships to generate universal informative representations that enhance the performance of various downstream geospatial tasks~\cite{zhang2024veccity,chen2024self}.
Different types of geovector data entities usually involve tailored pre-training methods to best reflect their spatial, temporal, and contextual features. We review the pre-training techniques for different types of geovector data.

\subsec{Point-based data.} 
Point-based geospatial data primarily represent specific locations, such as POIs and event occurrences. 
Existing works usually achieve pre-training on such data by constructing self-supervised learning tasks, such as location prediction, to capture the spatial and contextual relationships among these points and enhance general spatial representation learning.
For instance, GeoBERT~\cite{gao2022geobert} leverages a corpus of 17 million POIs across 30 Chinese cities to pre-train grid embeddings using masked POI type prediction, supporting multiple downstream tasks.
Besides, additional pre-training strategies are employed to improve the spatial representations. 
GeoVectors~\cite{tempelmeier2021geovectors} employs pre-trained word vectors from fastText~\footnote{https://fasttext.cc/} to encode OpenStreetMap entity tags, ensuring semantic meanings are effectively captured. Further, a graph-based approach using the DeepWalk algorithm is utilized to learn latent representations from the geographic relationships between entities.
TALE~\cite{wan2021pre} leverages abundant unlabeled trajectory data for self-supervised pre-training to learn location embedding vectors, incorporating universal location information to enhance downstream task performance.
G2PTL~\cite{wu2024g2ptl} pre-trains location representations using extensive geographic knowledge and spatial topology information, evaluating its performance across multiple GIS downstream tasks.


\subsec{Polyline-based data.} 
Polyline-based data describe linear entities (\eg road segments), representing the connectivity and sequential nature of elements. 
Pre-training on polyline data aims to understand and encode these inherent properties to improve tasks like navigation, route optimization, and traffic prediction.
One widely used pre-training strategy is polyline prediction that the model predicts the next point or segment in a polyline, thereby learning connected and sequential patterns. 
For example, Hier~\cite{shimizu2020enabling} pre-trains its model using real-world trajectory data through next-place prediction tasks, enabling finer-grained place embeddings. 
Toast~\cite{chen2021robust}, pre-trained with a traffic context-aware skip-gram module and a trajectory-enhanced Transformer module, effectively captures traffic patterns and travel semantics to enhance road network representation learning, applying to four downstream tasks.
Another prominent method is augmented polyline contrastive learning, which creates multiple augmented versions of an original polyline using data augmentation techniques. 
For instance, Mao et al.~\cite{mao2022jointly} proposes a pre-trained model that leverages contrastive learning to jointly learn road network and trajectory representations by maximizing mutual information between them. 
%
%

\subsec{Polygon-based data.} 
Polygon-based data involve area entities defined by closed boundaries, such as administrative regions, land parcels, or building footprints. Pre-training for polygon data focuses on learning general representations that capture spatial hierarchies and area-based attributes. 
One prominent approach is to encode geographic contextual information and complex interactions among spatial entities within a region. 
For example, GMEL\cite{liu2020learning} aims to pre-train effective embeddings for urban geographic units by encoding their geographic contextual information, which improves performance in tasks such as socio-economic prediction and urban function classification. 
Extending this idea, GeoHG\cite{zou2024learning} introduces an effective heterogeneous graph structure to learn comprehensive region embeddings. By leveraging the relationships between different types of spatial entities within a region, GeoHG supports various downstream tasks and enhances the richness of the learned representations.
Another method is region contrastive learning, which leverages the intuition that connected map entities in a relational network should have similar representations. 
RegionDCL~\cite{li2023urban} employs group-level and region-level contrastive learning strategies across varying region partitions, enabling adaptability to diverse spatial schemes and enhancing flexibility in different contexts.
In addition, recent research has focused on leveraging graph-based pre-training approaches for enhanced urban region representation.
GURPP~\cite{jin2024urban} constructs an urban region graph that integrates detailed spatial entity data and develops a subgraph-centric pre-training model. 
This model captures heterogeneous and transferable interaction patterns among entities, leading to improved urban region representations and better performance in tasks like urban planning and policy-making.

\subsubsection{Unimodal Adaptation} 
After pre-training, the learned representations of geovector data need to be adapted to specific downstream tasks to achieve optimal performance. 
This process often involves supervised learning where the pre-trained model serves as a foundation model, and additional layers or modules are trained using labeled data pertinent to the target task.
To name a few, AGB~\cite{muszynski2024fine} adapts a geospatial foundation model for above-ground biomass estimation by fine-tuning it with space-borne data tailored to the diverse eco-regions of Brazil.
Moreover, prompt tuning is introduced to enhance the adaptation on diverse downstream tasks. HREP~\cite{zhou2023heterogeneous} presents a heterogeneous region embedding module combined with prompt learning. By modeling the intricate interactions among diverse spatial entities within a region, this method dynamically adapts representations to task-specific requirements, enabling more nuanced and flexible applications across varying contexts.

\subsubsection{Cross-modal Adaptation} 
In addition to supervised fine-tuning, combining the geometric precision of geovector data with rich contextual information from other modalities, such as natural language, has proven beneficial. LLMs can capture more comprehensive features of geographic entities, leading to improved performance in tasks like location recommendation~\cite{ramrakhiyani2023zero}, semantic mapping~\cite{li2023geolm}, and urban planning~\cite{manvigeollm}.
For instance, Ramrakhiyani et al.~\cite{ramrakhiyani2023zero} evaluate the geographic knowledge learned by pre-trained language models using a probing dataset of 5,268 masked sentences derived from Wikidata triples. It probes 8 pre-trained language models through masked token prediction and text generation tasks.
Similarly, Maheshwary et al.~\cite{maheshwary2024pretraining} pre-train and fine-tune language models to determine whether two addresses represent the same physical building.

Despite the significant progress in geovector-based UFMs for point, polyline, and polygon data, handling large-scale geovector datasets poses challenges in terms of computational resources and processing time. 
Existing models often lack the scalability needed to effectively process extensive geospatial data like nationwide road networks or global POIs. 
Research into scalable and efficient algorithms or architectures is needed to empower pre-training and adaption for large-scale geovector datasets effectively. 
Furthermore, incorporating other modal data, such as satellite imagery and text descriptions, might enhance a more comprehensive understanding and representation for geospatial entities beyond only geovector data.

\subsection{Multimodal Models}
Due to the intricate nature of urban environments characterized by diverse data types, there is growing attention on developing multimodal UFMs to handle urban tasks. By integrating these disparate data modalities, multimodal UFMs aim to achieve a more holistic understanding of urban dynamics, largely enhancing various urban applications and setting the stage for urban general intelligence.
This section reviews current studies on building multimodal UFMs through multimodal pre-training and adaptation techniques.

\subsubsection{Multimodal Pre-training}
To develop a comprehensive understanding of urban environments, multimodal UFMs aim to systematically pre-train models on extensive multimodal datasets containing urban-specific information such as textual, visual, and geographic data. 

Among the existing studies, a considerable part of works focus on learning universal representations of urban entities, \eg~regions, roads, and POIs, which can support a variety of downstream tasks.
UrbanCLIP~\cite{yan2023urban} is a framework that applies LLMs to enhance multimodal urban region profiling. It leverages LLMs to generate textual descriptions for satellite images and employs contrastive learning as well as language modeling loss as supervision for urban visual representation learning.
Moreover, ReFound~\cite{xiao2024refound} leverages language, visual, and visual-language foundation models for urban region understanding. It learns in-domain knowledge from multi-modal geospatial data while distilling general knowledge from established foundation models, empowering the model with both general and domain knowledge. The generated region representations can be used for diverse urban analysis tasks like population prediction and commercial activity assessment.
Additionally, MM-Path~\cite{xu2024mm} presents a multi-modal and multi-granularity path representation learning framework that pre-trains road embeddings through contrastive learning by utilizing both road network-based paths and image-based paths. It effectively adapts to large-scale, real-world datasets across downstream tasks, including travel time estimation and path ranking.
Further, CityFM~\cite{balsebre2023city} is a framework that pre-trains universal representations of geospatial entities in a self-supervised manner using spatial, visual, and textual data from OpenStreetMap. It designs three contrastive objectives using nodes, ways, and relations to learn multimodal representations of geospatial entities and demonstrate their effectiveness in downstream tasks such as traffic speed inference and building functionality classification.

Beyond multimodal representation learning for urban entities, \rev{BIGCity~\cite{yu2025bigcity} develops a universal spatiotemporal foundation model that unifies trajectory data and traffic-state data through a shared ST-unit representation and a GPT-based prompt-driven framework. A two-stage training paradigm, consisting of masked reconstruction pre-training and multi-task prompt tuning, enables it to support eight heterogeneous tasks across both modalities without task-specific fine-tuning.}
Moreover, AllSpark~\cite{shao2023allspark} introduces the Language as Reference Framework (LaRF) to develop a multimodal spatio-temporal model that amalgamates thirteen distinct modalities into a single framework. LaRF's concept centers around correlating abstract concepts from each modality with language, facilitating a collective interpretation within the language-based unified representation space. 
In addition, TengYun~\cite{zhao2023parallel} is a transportation foundation model developed for TransVerse, focusing on cognition, operation, management, and control. It combines pre-training with prompt tuning for training, starting with pre-training on cloud servers using the Transformer framework and deep reinforcement learning to process historical multi-modal data. 
During prompt tuning, edge servers customize prompts for different tasks and strategies by treating training as a self-supervised regression problem.

\subsubsection{Multimodal Adaptation}
Another line of work explores introducing foundation models from other domains, \eg LLMs, to address urban challenges. The adaptation approaches of these works primarily include prompt engineering and model fine-tuning.

\subsec{Prompt engineering.}
The works using prompt engineering usually combine foundation models with domain-specific tools like Geographic Information Systems~(GIS) and external databases. 
TrafficGPT~\cite{zhang2023trafficgpt} is a framework for urban traffic management that integrates ChatGPT and various external tools. It involves prompt management, task understanding and planning through LLMs, and leverages LLMs to invoke available tools for specific executions~(\eg~database retrieval and analysis, data visualization, and system optimization). TrafficGPT also concludes with dialogue memory storage to enhance its conversational context for future interactions.
GeoGPT~\cite{zhang2023geogpt} is a framework integrating LLMs with mature GIS tools to assist non-professionals in solving complex geospatial tasks using natural language. It leverages LLMs to understand user demands and conducts automated data collection, processing, and analysis in GIS, enabling users to perform tasks like spatial queries and facility siting. 
Besides, Zhou et al. \cite{zhou2024large} apply LLMs to simulate planners and residents for participatory urban planning. This work presents a planning workflow where the planner proposes an initial land-use plan based on the image map and textual description of the region, and this plan is then revised through discussion with residents.

\subsec{Model fine-tuning.}
Beyond directly invoking the off-the-shelf foundation models, some studies also introduce a fine-tuning process for better adapting other domain foundation models to urban applications. 
TransGPT-MM~\cite{wang2024transgpt} is fine-tuned from the general multi-modal LLM, VisualGLM-6B~\cite{ding2021cogview}, to cater to the transportation domain specifically. This adaptation involves fine-tuning LLM on multi-modal domain-specific datasets derived from driving tests, traffic signs, and landmarks, showcasing powerful capabilities in traffic-related analysis and modeling tasks.
Xu et al. \cite{xu2023urban} integrate LLMs into urban systems to tackle urban challenges, using continued pre-training to infuse the CityGPT model with urban knowledge. It combines general texts, domain-specific data (\eg~urban knowledge graphs and geographic data), and task-solving processes, and then employs supervised fine-tuning to tailor CityGPT more closely to specific urban tasks and human preferences.
GeoReasoner~\cite{li2024georeasoner} is designed for geo-localization by fine-tuning the large vision-language model through a two-stage process. First, reasoning tuning is conducted using textual clues paired with street-view images to enhance the model's inference capabilities. Second, location tuning refines the model using curated high-locatability images, integrating visual and textual data for precise geo-localization.
Furthermore, VELMA~\cite{schumann2023velma} achieves an embodied LLM-based agent for vision and language navigation in the street view. It integrates visual observations from street view through CLIP for landmark recognition and uses a verbalization pipeline to translate visual information and navigation instructions into textual prompts for decision-making. It fine-tunes LLM on urban domain training instances using LoRA~\cite{hu2021lora}, which adapts attention projections in LLM.

Despite these works proactively advancing UFMs to handle multimodal urban data, they are primarily limited to integrating a restricted range of urban data types, particularly focusing on language and vision modalities, or applying to only a narrow scope of urban domains. Furthermore, the potential for spatio-temporal reasoning and the privacy and security issues in UFMs' development remain largely unexplored.
This reveals significant opportunities for creating more versatile, practical, and trustworthy UFMs.

\subsection{Other Models}
Except for the aforementioned categories, there exist some studies exploring the foundation models in other urban scenarios, such as tabular-related tasks~\cite{wen2024supervised}, decision making~\cite{lai2023large} and simulation~\cite{wang2023transworldng, da2024open}.
Specifically, in the field of intelligent transportation have emerged some studies on how to leverage the capacities of LLMs for complicated traffic control tasks~\cite{villarreal2023can,da2023llm,lai2023large}.
For example, 
PromptGAT~\cite{da2023llm} uses LLMs to generate human knowledge, that helps the neural networks understand realistic cases (\eg weather conditions) in traffic signal control tasks.
In addition, some preliminary efforts have been devoted to building a foundation model in transportation system modeling~\cite{wang2023building,wang2023transworldng}, which leverages the graph transformer architecture to dynamically explore the states of massive participants and their interactions.
Further, Open-TI~\cite{da2024open} integrates multiple traffic simulation platforms like SUMO and CityFlow. Augmented by LLMs, it supports complex simulation tasks such as map data processing from scratch, traffic signal control, and demand optimization.


\begin{figure*}[tb]
\centering
\includegraphics[width=1\columnwidth]{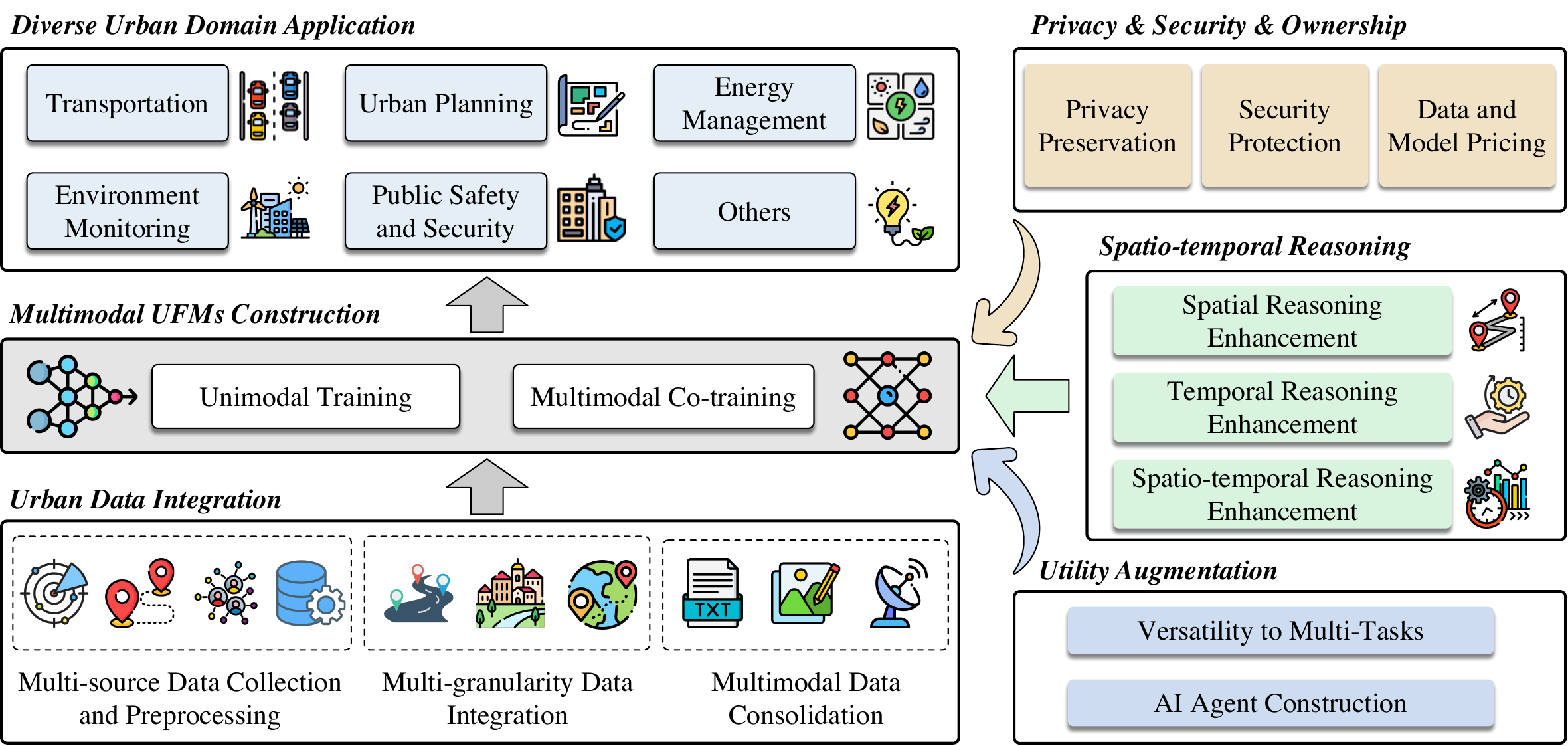}
\caption{\rev{A prospective framework for building versatile UFMs.}}
\label{fig:framework}
\end{figure*}

\section{Prospects of Urban Foundation Models} \label{solution}
While there have been burgeoning endeavors to develop UFMs, these attempts have predominantly focused on integrating a narrow range of urban data types and applying them to a limited set of urban tasks and contexts. The potential of UFMs for spatio-temporal reasoning, the versatility to multi-tasks, interaction with users and physical world, and the privacy, security, ownership issues when building UFMs remains largely untapped. 

Ideally, UFMs should be capable of processing multi-source, multi-granularity, and multi-modal urban data and be adaptable to a broad spectrum of urban tasks and domains. Additionally, they should possess intelligent spatio-temporal reasoning abilities to grasp and interpret the intricate dynamics and interconnections within urban environments. This should be achieved without compromising data privacy and model security while ensuring seamless interaction with the external environment, and enabling feedback-driven self-improvement.
Therefore, we present a prospective framework to overcome current obstacles and towards building versatile UFMs. The entire framework is illustrated in \figref{fig:framework}, which includes major components: urban data integration, multimodal UFMs construction, spatio-temporal reasoning, utility augmentation, and privacy, security, ownership. 

\subsection{Urban Data Integration} 
\subsubsection{Multi-source Data Collection and Preprocessing}
The standard procedure of building UFMs begins with multi-source urban data collection and preprocessing.

\subsec{Collect multi-source urban data.} The collection of urban data requires the aggregation of multiple data sources and diverse data types to form a comprehensive multimodal urban dataset~\cite{rathore2018exploiting}. The multi-source multimodal urban data may include but is not limited to geo-text data, social media data, street view images, remote sensing data, trajectory data, time series data, and geovector data~\cite{mai2023opportunities,luo2020spatial}. The granularity of these data is diverse, ranging from city-level~(\eg~weather monitoring data) to individual-level~(\eg~personal trajectory), from yearly records to real-time measurements.
The data collection process typically involves extracting from databases, crawling from the web, gathering from various deployed sensors, crowdsourcing~\cite{alvear2018crowdsensing,xin2021out}, \etc

\subsec{Preprocess multi-source urban data.} 
This stage, typically involving data cleaning, standardizing, normalizing, and annotating as needed, is particularly critical to ensure the usability and effectiveness of the data in foundation models training and analysis~\cite{chen2022spatio,zha2023data}. Data cleaning identifies and corrects (or removes) errors, noise, and inconsistencies in the data to improve its quality. Data standardization aligns data formats and structures to achieve consistency and ensure interoperability across multiple sources~\cite{gal2019data}.
Data normalization ensures that data from various sources are on a comparable scale. Data annotation is required by supervised learning of models for a better understanding of specific urban tasks and scenarios. This might involve tagging images, categorizing text, marking events in time series, \etc

\subsubsection{Multi-granularity Data Integration}
Multi-granularity data integration aims to employ suitable data processing methods and GIS tools to realize a meticulous blend and correlation for multi-granularity urban data. We present the data scaling, hierarchical data structuring,  and data cross-referencing to enable an effective integration of multi-granularity data.


\subsec{Data scaling.}
Urban data often originate from diverse sources with varying spatial and temporal resolutions. To integrate these datasets effectively, downscaling and upscaling techniques can be employed to harmonize their scales~\cite{ge2019principles}.
Downscaling refines coarse-resolution data to achieve higher spatial or temporal details, aligning it with finer-scale datasets. This is particularly useful when integrating global or regional data with local observations. For instance, statistical downscaling methods~\cite{wilby2013statistical} can adjust large-scale climate projections to reflect local urban climates more accurately. 
In contrast, upscaling aggregates high-resolution data to a coarser scale, facilitating integration with broader datasets. This approach is beneficial when detailed local measurements need to be incorporated into regional or global data. 

\subsec{Hierarchical data structuring.}
Organizing the data into a hierarchical structure is an effective strategy in dealing with the complexity and diversity of urban data~\cite{hadlak2010visualization}, as it allows for a more nuanced representation that closely mirrors the real urban landscape. At the top of this hierarchy would be city-level data, encompassing broad metrics and trends that define the city as a whole. This could include overarching demographic statistics, city-wide economic indicators, or general environmental data. Beneath this top layer, the hierarchy would then break down into more localized data sets, \eg districts, neighborhoods, and individuals. Each layer offers a finer resolution of data, capturing the unique characteristics and dynamics of smaller areas. For instance, district-level data might focus on regional traffic patterns or specific economic activities, while neighborhood-level data could delve into local public services, housing trends, and community-specific issues. At the most granular level, individual data may include detailed individual trajectories and other mobile records. 
Such a structuring aids in systematically analyzing data at each level and understanding their intricate relationships and interactions.

\subsec{Data cross-referencing.}
Cross-referencing multi-granularity urban data is a complex process that is crucial for gaining comprehensive insights into urban environments.
It involves identifying, aligning, and correlating relevant data from various data granularity, such as correlating the micro-level data, \eg~driver trajectories, individual emissions, or local crime reports, with macro-level data, \eg~city-wide traffic measurements, general air pollution indices, or overarching economic indicators, thereby, providing valuable insights from different urban data resolution for each other.
To achieve cross-referencing for multi-granularity urban data, a systematic approach necessitates begins with the standardization of varied data sources~\cite{gal2019data}. 
Advanced GIS tools are usually employed to enable the overlay of diverse datasets on a common spatial grid. 
Once aligned, analytical techniques like statistical correlation analysis, spatial analysis, and machine learning algorithms are applied to identify patterns and relationships within the integrated data~\cite{shekhar2011identifying,atluri2018spatio}. This phase is pivotal in extracting meaningful insights from the complex interplay of varied urban parameters. Finally, the insights can be presented through data visualization tools~\cite{zhong2012spatiotemporal}, which translate the complex data relationships into understandable and actionable formats, facilitating the UFMs design for researchers.


\subsubsection{Multimodal Data Consolidation} \quad
\label{mmdata}
Data from different modalities are heterogeneous but are often connected due to shared complementary information. Multimodal data consolidation is crucial in developing UFMs as it provides the foundation for integrating diverse data types to create a comprehensive understanding of urban environments from multiple data perspectives. This holistic view is fundamental to enhancing UFMs with robust generalization and adaptation capabilities across a wide range of urban tasks and domains. This involves multimodal data representation and alignment.

\subsec{Data representation.}
To harmonize heterogeneous data~(\eg text, image, trajectory, time series) and facilitate their consolidation, the first step is to encode the diverse urban data modalities into a latent representation space~\cite{liang2024foundations}. This can be achieved by respectively developing embedding layers for each data modality~\cite{baltruvsaitis2018multimodal} or utilizing the corresponding established unimodal foundation model as an encoder~\cite{wu2023next}. For instance, text data is processed through Transformers or LLMs, and images through CNNs or VFMs. 

Representation fusion, coordination, or fission are usually adopted to create a unified representation that captures diverse characteristics and interactions across modalities~\cite{liang2024foundations}. Fusion integrates information from multiple modalities into a joint representation, often using additive, multiplicative, attention-based, or tensor-based interactions to reflect shared and unique modality features~\cite{zhao2024deep}. Coordination aligns and refines representations by mapping different modalities into a shared space, either through strong alignment~\cite{radford2021learning}, which forces high similarity for semantically related elements, or partial coordination~\cite{liang2024foundations}, which preserves general relationships. Lastly, fission decomposes multimodal data into modality-specific and shared components, often using techniques like disentangled representations~\cite{tsai2019learning} or clustering~\cite{hu2019deep}, to capture fine-grained modality interactions while preserving their individual structures. These methods jointly enable coherent and flexible representations that support diverse multimodal tasks.

\subsec{Data alignment.} 
Alignment aims to establish meaningful connections between elements across different modalities, ensuring they are synchronously represented for integrated processing. There are three primary categories of data alignment methods for multimodal learning~\cite{liang2024foundations}.

Discrete alignment focuses on linking specific, well-defined elements across modalities. Local alignment connects clear segments, like words in text with object regions in images, by employing methods such as contrastive learning, which matches representations of similar concepts from different modalities~\cite{radford2021learning}. When explicit pairings are unavailable, global alignment can be employed to optimize the relationships between all elements across both modalities, using techniques like Optimal Transport to minimize divergence between modality spaces~\cite{chen2020graph}.

Continuous alignment handles data that lacks clear segmentation, requiring synchronization of continuous streams across different modalities. It is suitable for aligning continuous sensory data like trajectory and time series. For example, aligning the continuous streams of traffic cameras and GPS devices to monitor real-time traffic conditions, and synchronizing the weather station measurements and satellite imagery to track dynamic environmental conditions. 
Techniques include continuous warping, which forms a bridge between two representation spaces by aligning modalities through adversarial training~\cite{munro2020multi} or dynamic time warping~\cite{trigeorgis2017deep}, effectively transforming one representation to match another. Additionally, segmentation methods divide these high-dimensional signals into meaningful units by identifying semantic boundaries, such as temporal segmentation~\cite{yuan2008speaker} and clustering~\cite{sun2019videobert}. These methods are essential for tasks where modalities exhibit temporal or spatial ambiguity, enabling structured alignment and improving cross-modal understanding.

Contextualized alignment goes beyond basic pairing by automatically learning how elements from one modality interact with all related elements from other modalities, integrating them into a cohesive representation. Key approaches include joint undirected alignment, which focuses on capturing symmetrical connections between modality elements, often using methods like multimodal transformers to incorporate all modality elements in alignment~\cite{li2019visualbert}. Another approach is cross-modal directed alignment, where elements from one modality are directed toward another, leveraging mechanisms like query-key-value attention to model asymmetric connections~\cite{tsai2019multimodal}. Lastly, graphical alignment generalizes beyond sequential patterns, using graph networks to structure modality interactions, allowing arbitrary connections between elements to model relationships across modalities effectively~\cite{yang2021mtag}.

\subsection{Multimodal UFMs Construction}
To build versatile UFMs that can handle diverse urban tasks and domains, the key is to augment UFMs' capabilities to understand multimodal urban data. 
However, directly building a unified UFM handling all urban data modalities in end-to-end manners would be extremely challenging because of the significant heterogeneity and distinct granularity from diverse data types and sources.
A more plausible way towards multimodal UFMs is to further decompose the goal and incorporate both unimodal training and multimodal co-training. This not only offers flexibility and potential in integrating an unlimited range of modalities but also allows for more tailored approaches to accommodate the heterogeneity of different modalities.


\rev{
\subsubsection{Unimodal Training} 
This refers to training unimodal UFMs for various urban data modalities separately, which can be achieved by training foundation models from scratch or adapting established foundation models.

\subsec{Training foundation models from scratch.} 

When a data modality offers abundant, high-quality signals but lacks any suitable pretrained models, building a foundation model from scratch becomes a viable—though resource-intensive—approach. In such cases, self-supervised learning provides a natural pathway for large-scale pretraining. Typical strategies include next-step forecasting for time series, next-location prediction for mobility trajectories, or contrasting geovector entities like POIs and road networks based on their contextual similarities~\cite{dasdecoder, yuan2024unist, t2vec2018, lightpath2023, balsebre2023city}. These objectives allow the model to autonomously uncover temporal dependencies, movement regularities, spatial correlations, and functional similarities underlying urban big data. Additionally, supervised learning can be applied when labeled data is available, further enhancing model's specialization on domain-specific tasks and urban contexts~\cite{hmtrl2022}.

When building foundation models from scratch, a central challenge is the significant computational burden associated with large-scale self-supervised pretraining. To address this, several techniques have emerged to make the process more lightweight and resource-efficient. 
One strategy is to adopt modular or lightweight architectures, such as MoE~\cite{shazeer2017outrageously,shi2024timemoe} or linear-complexity architectures~\cite{gu2024mamba}, allowing the model to scale in capacity without significantly increasing computation overhead. 
Another direction is the use of compact architecture design tailored to the structural properties of urban data examples, such as multi-scale mixer~\cite{han2025scalable} and hierarchical transformers~\cite{bi2023accurate}, to efficiently process the multi-resolution spatio-temporal inputs.
Additionally, efficient training paradigms such as curriculum pretraining~\cite{wang2020curriculum} and knowledge distillation~\cite{he2022knowledge} can substantially cut training time by first learning coarse-grained patterns and then refining only the most informative components. 
From an optimization perspective, techniques like mixed-precision training, smart model scales and initialization, and selective token or instance training, also substantially help reduce memory footprint and computation demand~\cite{wan2024efficient}. 
Together, these strategies make it possible to build more compact yet expressive foundation models that retain strong representational power while remaining practical for the compute-constrained settings typical of urban research and deployment.

\subsec{Adapting established foundation models.} 
For data modalities already with general foundation models, such as LLMs and VFMs, adapting these pre-existing foundation models to the specific context of urban data is typically more computationally efficient and economical than establishing one from scratch. The adaptation is usually achieved by: (1)~Prompt methods~\cite{liu2023pre}, including prompt engineering and prompt tuning. Prompt engineering typically uses well-designed prompts to harness the power of pre-existing foundation models for the urban environments without changing foundation models' parameters~\cite{ning2024urbankgent,wang2023samrs,auxmoblcast2022}; Prompt tuning adjusts a small set of parameters associated with specially designed input prompts, rather than the entire model~\cite{lester2021power}. These prompts act as adaptable guides, helping the model to focus and improve on specific tasks like understanding specific urban contexts or generating thematic responses~\cite{zhang2023promptst,yuan2024unist,zhao2023parallel}. Prompt methods effectively harness the large model's pre-trained knowledge base, allowing for efficient and task-specific adaption without extensive retraining; 
(2)~Model fine-tuning methods: the model's parameters are fine-tuned using a labeled dataset in a supervised manner, where the urban data has annotations that guide the model towards better performance in tasks like street image recognition or geo-text reasoning~\cite{vivanco2024geoclip,lai2023large}. Alternatively, it leverages the patterns inherent in the data to construct the self-supervised objectives without needing explicit labels, enhancing the model's understanding of urban context~\cite{li2023geolm,ansari2024chronos}. As tuning the entire large foundation model would be computationally expensive and could lead to catastrophic forgetting, current studies prefer to refine the model using parameter-efficient fine-tuning techniques~\cite{hu2021lora}.

For modalities that lack established foundation models, some proactive works attempt to align the urban data into the representation space of existing other modal foundation models~(\eg~LLMs, VFMs). This might be done by: (1) Directly prompting the foundation models like LLMs to understand urban data~\cite{llmmob2023};
(2)~Training lightweight encoders or decoders to transform the specific urban data modality into the representation space that is comprehensible by existing foundation models~\cite{li2024urbangpt,zhou2023one1_nips}; 
(3)~Utilizing model reprogramming techniques~\cite{chen2024model} to reformat the data to mimic the format that pre-trained foundation models expect~\cite{wang2024empowering,jin2024time}. 
}



\subsubsection{Multimodal Co-training} 
This aims to build cross-modal connections, interactions, and information transfers to enrich a comprehensive multimodal understanding of UFMs. 
Typically, each modality data is first embedded by the corresponding modality encoder, which might be instantilized by modality-specific embedding layers~\cite{baltruvsaitis2018multimodal} or the corresponding established unimodal foundation models~\cite{wu2023next}. 
After that, a learnable interface is introduced to bridge the gap between modalities, aiming to project the information from different modalities into a unified semantic space, \eg~a shared representation space or a specific modality space like text, achieving multimodal alignment~\cite{yin2023survey}. During the process, the pre-trained foundation models might keep frozen and only training the learnable interfaces~\cite{liu2024visual, li2024llava}, aiming to align different modalities without losing pre-trained knowledge. 
Alternatively, some pre-trained parameters could be fine-tuned to enable a more flexible multimodal alignment~\cite{wang2024visionllm,bai2023qwen,ye2023mplug}. The multimodal co-training may involve established foundation models or some new modules to be pre-trained, depending on specific domains and applications~\cite{xiao2024refound,yan2023urban}. 

In the co-training process, several techniques are usually used to align different modalities into a unified semantic representation space that encapsulates the multifaceted nature of urban environments, enabling more comprehensive understanding and insights.
(1)~Cross-modal mapping: Develop mapping functions that
transform the output of one modality to the space of
another~\cite{liu2024visual}. For instance, mapping the feature space of VFMs
to the feature space of LLMs. This is usually achieved
through training on paired data, \eg~image-caption pairs;
(2)~Regularization techniques: Employ regularization terms like contrastive loss~\cite{radford2021learning} or triplet loss~\cite{semedo2020adaptive} to align the representation of different modalities. These losses encourage representations of similar items across modalities to be close in the latent space while pushing dissimilar items apart; 
(3)~Canonical Correlation Analysis (CCA): Apply CCA to transform different modality representations into a shared subspace where their correlations are maximized, preserving inter-modal relationships and enhancing correlations between modalities~\cite{sun2020learning}.



\subsection{Spatio-temporal Reasoning}
Spatio-temporal reasoning refers to the ability to understand and analyze both the spatial (geographical and locational) and temporal (time-related) aspects of urban data in an integrated manner. It is crucial to empower UFMs to equip more comprehensive understandings and accurate predictions for urban dynamics, facilitating effective planning and management of complex urban environments.

\subsubsection{Spatial Reasoning Enhancement}
Spatial reasoning is a critical component for enabling UFMs to navigate and interpret the intricate spatial relationships that define urban environments. This section presents two techniques: universal geo-position embeddings and symbolic reasoning, highlighting the potential to empower spatial representation and reasoning.

\subsec{Universal geo-position embeddings.}
Inspired by the success of word embeddings~(\eg~word2vec~\cite{mikolov2013efficient}) in LLMs, it is intuitive to learn universal geo-position embeddings for inter-correlated locations across the Earth~\cite{mai2020multi}. 
Specifically, the spatial similarity and correlations between locations can be measured based on multiple factors such as geographical distance, transportation accessibility, inter-regional flow, and regional functionalities. 
Then universal geo-position embeddings are pre-trained by using self-supervised methods like contrastive learning or context prediction (analogous to word2vec). Once pre-trained, these geo-position embeddings can serve as universal spatial representations across various modalities, enabling cross-modal geospatial alignment and enhancing spatial reasoning capabilities in all data types~\cite{mai2023opportunities}. Additionally, the positional encodings for objects, such as relative coordinates or patch positions, are useful in enhancing visual-spatial reasoning~\cite{liu2023visual}.

\subsec{Symbolic reasoning.}
Spatial reasoning can be enhanced by representing spatial relationships with concise symbolic notations during intermediate reasoning steps. This simplifies the representation and reduces the ambiguity of complex environments, focusing the model's attention on essential spatial information~\cite{hu2024chain}. Furthermore, symbolic training enhances spatial reasoning by integrating explicit logical rules into the training process~\cite{premsri2024neuro}. This can be achieved by translating logical constraints into differentiable soft logic, which is then added to the loss function, guiding the model to follow spatial logic rules. By enforcing consistency across a chain of reasoning steps based on spatial rules, the model can learn to handle complex spatial questions.

\subsubsection{Temporal Reasoning Enhancement}
Temporal reasoning is a vital capability for UFMs to understand and predict the dynamic changes and causality in urban systems over time. Enhancing temporal reasoning might involve leveraging external knowledge, integrating logical rules, and incorporating time encoding techniques.

\subsec{External knowledge.}
Integrating external knowledge sources, like knowledge graphs~(\eg~ConceptNet~\cite{speer2017conceptnet}, $\text{ATOMIC}^{20}_{20}$~\cite{hwang2021comet}), can supply additional context about event order and duration~\cite{yuan2024back}. It helps mitigate UFMs' reliance on biased or incomplete data by incorporating established event patterns and causal relationships, which also guide UFMs to better understand rare or novel events by leveraging prior information, enabling more robust predictions and reasoning across varied urban scenarios~\cite{ning2024urbankgent}.

\subsec{Logical reasoning.} UFMs can incorporate explicit rules and logic to systematically represent relationships between temporal events. For instance, probabilistic logic can enforce constraints like ``an event lasting a year must recur less than annually''~\cite{cai2023self}. By integrating logical rules into reasoning, the models are allowed to generalize temporal rules rather than rely solely on data-driven associations, thus supporting more reliable event ordering and duration estimation~\cite{cai2022mitigating,yuan2024back}. 

\subsec{Time encoding.} Temporal reasoning can be enhanced by incorporating time encoding at various temporal granularities (\eg~year, month, day, hour, minute, second)~\cite{shankaranarayana2021attention}. Additionally, continuous time can be encoded using learnable trigonometric functions to capture periodic features~\cite{zhang2023irregular}. Time encoding can enhance temporal understanding and reasoning for different data modalities and tasks, such as spatio-temporal prediction~\cite{zhang2023irregular,liu2023spatio} and text generation~\cite{cao2022time}.

\subsubsection{Spatio-temporal Reasoning Enhancement}
This section explores strategies for strengthening integrated spatio-temporal reasoning by exposing UFMs to reasoning-augmented datasets and tasks while improving deliberate reasoning capabilities. 

\subsec{Learning on reasoning-augmented datasets and tasks.}
The spatio-temporal reasoning capabilities can be augmented by adapting UFMs to the reasoning datasets or tasks rich in spatial and temporal dynamic features, \eg~spatio-temporal QAs~\cite{li2025stbench}, complex travel planning tasks~\cite{xie2024travelplanner}, locatable street-view images paired with human-inference knowledge~\cite{li2024georeasoner}, and the complex event with temporal reasoning paths~\cite{yuan2024back}. These datasets or tasks are designed to cover multi-dimensional spatial and temporal information in the real world and to ensure the diversity and representativeness of scenarios. They can be used for instruction tuning~\cite{feng2024citygpt,yuan2024back}, in-context learning~\cite{li2025stbench}, tool learning~\cite{xie2024travelplanner}, and reasoning tuning~\cite{li2024georeasoner} of UFMs to strengthen the spatial-temporal reasoning capabilities.

\subsec{Learning deliberate reasoning.}
Since spatio-temporal reasoning is an aspect of general reasoning ability, enhancing the model's overall reasoning capacity should also strengthen its spatio-temporal reasoning skills. 
The success of OpenAI o1\footnote{\url{https://openai.com/index/introducing-openai-o1-preview/}} has shown that engaging in a more thorough internal thought in reasoning processes can significantly enhance the reasoning capabilities of LLMs, which suggests that increasing deliberate thinking could similarly improve spatio-temporal reasoning. 
To cultivate deliberate reasoning, strategies such as Chain of Thought~\cite{wei2022chain}, Tree of Thought~\cite{yao2024tree}, bootstrapping reasoning~\cite{zelikman2022star}, internal deliberation before response~\cite{zelikman2024quiet}, retrieval-augmented self-critique~\cite{asai2023self}, allocating additional computational resources during inference~\cite{snell2024scaling}, and reasoning with reinforced fine-tuning~\cite{trung2024reft} can be utilized to substantially enhance LLM-based models reasoning.


\subsection{Utility Augmentation}
The utility of UFMs lies in their versatility to handle a wide range of tasks across diverse urban domains, seamlessly interact with users and the physical world, continuously evolve through ongoing updates and learning, effectively leverage available external resources, \etc

\subsubsection{Versatility to Multi-tasks}
UFMs' versatility to handle a wide range of tasks is essential for their utility in urban systems. This versatility reduces reliance on specialized models for each task while enhancing adaptability to new tasks and evolving urban demands, maximizing their applicability and impact. 
This section explores several techniques for enhancing UFMs' versatility to multi-tasks.

\subsec{Prompt learning.}
This method involves designing prompts that guide models toward desired outputs. By crafting effective prompts, models can be steered to generate relevant and accurate responses, facilitating efficient adaptation to various tasks without extensive retraining~\cite{liu2023pre}. Prompt learning has been employed to enhance the model's versatility to urban tasks, such as spatio-temporal prediction~\cite{yuan2024unist,zhang2023promptst},  region representation~\cite{jin2024urban}, and virtual urban renewal~\cite{hu2024ursimulator}.

\subsec{Instruction tuning.}
In this approach, models are trained to follow explicit instructions, improving their ability to understand and execute specific tasks. This method enhances zero-shot task generalization, allowing models to perform unseen tasks by interpreting and adhering to given directives~\cite{zhang2023instruction}. Instruction tuning has been used to enhance GIS-specific LLMs~\cite{zhang2024bb}, zero-shot spatio-temporal prediction~\cite{li2024urbangpt} and human mobility prediction~\cite{tang2024instruction}.

\subsec{In-context learning.}
This technique enables models to perform tasks by conditioning on input prompts that include examples or instructions, without explicit parameter updates. By leveraging patterns within the provided context, models can generalize to new tasks, enhancing adaptability across diverse applications~\cite{dong2022survey}. In urban scenarios, in-context learning has been applied to enhance the forecasting of human movement~\cite{llmmob2023} and time series~\cite{lu2024context}.

\subsec{Multi-task learning.}
This technique involves training models on multiple tasks simultaneously, enabling them to learn shared representations and improve performance across tasks~\cite{zhang2021survey}. By leveraging commonalities among tasks, models can generalize better and become more versatile in handling diverse challenges, such as multi-task visual understanding~\cite{luo2024delving} and spatio-temporal prediction~\cite{yi2024get}.

\subsec{Transfer learning.}
This method allows models to apply knowledge gained from one task to improve performance on a related task. By transferring learned features or representations, models can adapt more quickly to new tasks, reducing the need for extensive training data and computational resources~\cite{zhuang2020comprehensive}. Current research usually applies transfer learning for the knowledge transfer between different spaces, \eg~from data-rich cities to the data-scarce city~\cite{jin2022selective, zhang2024meta}.

\subsubsection{AI Agents Construction}
Building urban AI agents that leverage LLMs offers transformative potential for dynamically interacting with users and the physical world, intelligently leveraging external resources, and facilitating complex decision-making processes~\cite{wang2023survey}. 

\subsec{Interaction with users and physical world.}
LLMs facilitate intuitive conversational interfaces, allowing users to engage with AI agents using natural language and even multimodal queries~\cite{xie2024large}. This capability enables agents to comprehend complex user queries, retrieve relevant data, and provide transparent and explainable insights~\cite{lai2023large, zhang2023trafficgpt, zheng2023chatgpt}.
Additionally, LLMs can be also integrated with Internet of Things (IoT) sensors to interpret and analyze sensor data, effectively bridging the gap between digital intelligence and physical environments, and allowing AI agents to comprehend and respond to real-world scenarios~\cite{xu2024penetrative,guan2024citygpt}.

\subsec{Continual update and learning.}
Urban environments are continuously evolving; therefore, urban AI agents should support continual updates and learning mechanisms, which enable agents to adapt to new data patterns, seasonal shifts, and emergent urban trends. This might involve techniques such as 
(1)~Self-evolution~\cite{guo2024large}: Agents dynamically adjust their planning strategies or goals based on feedback from the environment or humans;
(2)~Retrieval augmentation~\cite{fan2024survey}: Retrieving and integrating stored memory that contains historical experience or external knowledge bases that are continually refreshed;
(3)~Continual learning~\cite{shi2024continual,wang2024comprehensive}: Update or adjust the model's parameters with new data to enhance its foundational or specialized knowledge; 
(4)~Online learning~\cite{hoi2021online}: Process data sequentially and update the model in real-time as new data arrives.

\subsec{Intelligent coordination for models and tools.}
The ability to invoke and coordinate specialized models and tools~(\eg~GIS tools) can largely enhance UFMs' profession and credibility in specific domains.
LLMs, with their advanced language understanding and reasoning capabilities, can serve as centralized orchestrators that manage and harmonize diverse specialized models and urban management tools~\cite{zhang2023trafficgpt}. Recent tool learning techniques~\cite{qu2024toolsurvey} empower the LLMs to extend its capabilities beyond its core functions by accessing external resources, making its outputs more accurate, comprehensive, and informative. This integration is particularly beneficial for complex queries where additional information or functions are required~\cite{qin2023toolllm}.

\subsec{Multi-agent cooperation.}
While a single AI agent may be limited by its capability, urban intelligence can benefit significantly from multi-agent cooperation, where multiple agents, each responsible for sub-tasks, work together to complete more complex urban tasks. 
The LLM-based agents are usually first created with specific roles and profiles, which enhance their expertise to perform designated sub-tasks in specific domains~\cite{lai2023large}.
The multi-agent systems are communicated using natural language and can be organized in layered, decentralized, centralized, or shared-message-pool structures to facilitate effective communication~\cite{guo2024large}. This allows agents to share information and strategies, and work collectively, thereby facilitating multi-agent cooperation.

\subsec{World simulation.}
World simulation empowers urban AI agents with a controlled environment to test and optimize decision-making strategies before deployment. 
The latest generative models, \eg~Earth-2\footnote{\url{https://www.nvidia.com/en-us/high-performance-computing/earth-2/}}, Sora\footnote{\url{https://openai.com/index/video-generation-models-as-world-simulators/}}, UrbanWorld~\cite{shang2024urbanworld}, and LLM-based world models~\cite{ge2024worldgpt,feng2024citybench,yan2024opencity}, open a new window to the nexus of the physical and cyber world.
This not only minimizes the potential real-world risks but also helps agents anticipate and prepare for rare or unprecedented situations.

\subsection{Privacy, Security, and Ownership}
Protecting sensitive data, ensuring model safety, and fostering equitable ownership are essential for the reliability, trustworthiness, and prosperous development of UFMs. This section discusses UFMs' privacy preservation, security protection, and data and model pricing.

\subsubsection{Privacy Preservation}
UFMs necessitate to be trained on vast multi-source urban data that are collected from different institutes or persons. However, these urban data often contain sensitive information, such as personal mobility patterns and social records, which can raise severe privacy concerns~\cite{gao2013trpf,shi2010prisense}. 
Traditional centralized training pools all data in one location, increasing the risk of breaches and misuse. Federated learning~\cite{yang2019federated} addresses this by enabling models to learn from decentralized data sources without sharing raw data, thereby protecting data privacy.
Advanced federated learning strategies are employed, \eg~model parallelism~\cite{yuan2022decentralized}, parameter-efficient training methods~\cite{lu2023fedclip,zhao2023fedprompt}, model pruning~\cite{jiang2022model}, and model compression~\cite{chen2022fedobd}, to ensure the learning process is feasible and more efficient for large foundation models.

In addition, it is essential to mitigate potential privacy breaches during user interactions with UFMs. Secure data protocols should be implemented to protect user information from being inadvertently revealed through model outputs or interaction logs~\cite{iqbal2024llm}. Anonymization, encryption, and differential privacy can prevent the inadvertent disclosure of sensitive information, ensuring user privacy throughout interaction processes~\cite{zhang2024privacyasst,du2023dp}.

Furthermore, the privacy of model parameters should be considered, as they may inadvertently capture and reveal private or sensitive data from training. Techniques such as parameter pruning~\cite{jiang2022model} and privacy-preserving model distillation~\cite{shao2024selective} can be adopted to reduce the risk of parameter leakage. These approaches, coupled with strong access control, contribute to the overall privacy resilience of UFMs.

\subsubsection{Security Protection}



In constructing UFMs, it is crucial to implement robust security measures to defend against malicious attacks, prevent the generation of harmful or misleading outputs, and ensure alignment with human preferences.

UFMs might be susceptible to multiple forms of malicious attacks, such as backdoor, poisoning, adversarial, and jailbreak attacks~\cite{yao2024survey}. The UFMs could incorporate protection strategies, such as data scrutiny~\cite{kurita2020weight}, adversarial training~\cite{chen2021pois,schwinn2023adversarial,liu2024adversarial}, unlearning~\cite{lu2024eraser}, and safety training~\cite{xu2024bag}, to bolster their defenses against these malicious attacks, thereby mitigating the risks in practical applications.

Another critical challenge for UFMs is ensuring their outputs are safe, factual, and free of harmful content. Hallucinated outputs, or responses that are factually incorrect or misleading, can be particularly detrimental when used in important decision-making or emergency management. Techniques such as safety alignment~\cite{ji2024beavertails}, self-refinement, and retrieval augmented generation~\cite{tonmoy2024comprehensive} can be applied to counter these issues.

For UFMs to effectively serve urban citizens, it is essential to align model outputs with human preferences, values, and social norms. This alignment process usually involves fine-tuning foundation models on datasets that reflect human preferences and integrating reinforcement learning techniques that reward outputs aligning with human feedback~\cite{wang2023aligning}. Regular feedback loops, sourced from urban citizens, public stakeholders, or city authorities, are important to help refine model responses to meet human expectations more accurately~\cite{wang2023towards}.

\subsubsection{Data and Model Pricing}
The success of UFMs depends on fair compensation mechanisms that encourage data sharing and model development. For data pricing, it is vital to implement a contribution measurement framework that quantifies each data provider’s input to the UFMs~\cite{wang2019measure}. Incentive mechanisms can encourage various stakeholders to share high-quality, diverse datasets, ultimately enhancing the UFM’s generalization and robustness~\cite{tu2022incentive}.

Similarly, model pricing strategies can be incorporated to incentivize the ongoing development and enhancement of UFMs~\cite{chen2019towards}. By establishing a value system that rewards model improvements based on utility, robustness, and policy alignment, UFMs can attract a broader community of developers and researchers. This pricing framework ensures sustainable, collaborative efforts in UFMs' development, fostering a balance between innovation and long-term operational success.

\rev{
\section{Benchmarks and Datasets} 
\label{dataset}

\begin{table*}[th]
\centering
\tiny
\setlength{\tabcolsep}{4.0pt}
\renewcommand{\arraystretch}{1.15}
\begin{tabular}{p{1.5cm} p{1.9cm} p{3.0cm} p{3.0cm} p{3.0cm}}
\hline
\textbf{Benchmark} & \textbf{Data Modalities} & \textbf{Evaluated Abilities} & \textbf{Tasks} & \textbf{Metrics} \\
\hline

\textbf{USTBench}~\cite{lai2025ustbench} &
Language + Time Series + Trajectory + Geovector &
Spatiotemporal understanding; forecasting; planning; reflection-with-feedback; downstream applications &
4 prediction tasks: next-POI prediction; traffic congestion forecasting; socio-economic indicator prediction; OD prediction.
5 decision-making tasks: traffic-signal control; POI placement; route planning; road planning; urban planning &
Process-level structured QA evaluation + Task-specific quantitative metrics (e.g., MAPE/RMSE/Accuracy depending on task) \\
\hline

\textbf{CityBench}~\cite{feng2024citybench} &
Vision + Language + Trajectory + Geovector &
Urban perception \& understanding; planning \& decision-making &
8 tasks: image geo-localization; geospatial prediction; infrastructure inference; GeoQA for city elements; mobility prediction; urban exploration; outdoor navigation; traffic signal control &
Task-specific quantitative metrics (e.g., Acc@1km/25km; $R^2$/RMSE; Recall@K/Top-1; success rate/avg steps/queue length) \\
\hline

\textbf{STBench}~\cite{li2025stbench} &
Language + Time Series + Trajectory + Geovector &
Knowledge comprehension; spatiotemporal reasoning; accurate computation; downstream applications &
15 tasks (4 dimensions): POI tasks; point tasks; trajectory tasks; region tasks (e.g., recognition/identification; relation reasoning; trajectory analysis/direction; anomaly detection/prediction/classification; flow prediction; navigation) &
QA accuracy + Task-specific quantitative metrics (e.g., absolute error for trajectory prediction; MAE/RMSE for flow prediction) \\
\hline

\textbf{STARK}~\cite{quan2025benchmarking} &
Language + Time Series + Trajectory &
State estimation; spatiotemporal reasoning; world-knowledge-aware reasoning &
Field-variable prediction; localization/tracking; spatiotemporal relationship inference; landmark-/intent-aware navigation-style tasks &
Challenge-level success/accuracy \\
\hline

\textbf{UrBench}~\cite{zhou2025urbench} &
Vision + Language &
Geo-localization; scene understanding \& reasoning; object understanding; cross-view consistency &
14 QA tasks across 4 dimensions: geo-localization, scene understanding, scene reasoning, object understanding &
QA accuracy\\
\hline

\textbf{CityEQA}~\cite{zhao-etal-2025-cityeqa} &
Vision + Language &
Embodied active exploration; long-horizon planning; spatial reasoning; open-vocabulary embodied QA &
1.4k human-annotated QA tasks (6 categories) in the EmbodiedCity simulator &
QA accuracy + Exploration-efficiency \\
\hline

\textbf{UrbanVideo-Bench}~\cite{zhao-etal-2025-urbanvideo} &
Vision + Language &
Urban video understanding: recall; perception; reasoning; navigation \& planning &
5.2k MCQs covering 16 tasks and 4 aspects: recall, perception, reasoning, navigation) &
MCQ accuracy \\
\hline

\textbf{CityLens}~\cite{liu2025citylens} &
Vision + Language & Socioeconomic indicators prediction &
Prediction tasks across economy, education, crime, transport, health, environment &
Coefficient of determination~($R^2$) and normalized RMSE (nRMSE) \\
\hline

\textbf{LLM-UrbanPlan}~\cite{zhao2025urban} &
Language &
urban planning documentation; examinations; routine data analysis; AI algorithm support; thesis writing &
556 tasks across urban planning documentation; examinations; routine data analysis; AI algorithm support; thesis writing &
Mixed automatic and human evaluation (e.g., human-rated accuracy/clarity/completeness/conciseness for explanations; correctness/syntax/clarity for code) \\
\hline

\textbf{Travel-Planner}~\cite{xie2024travelplanner} &
Language &
long-horizon planning; constraint satisfaction; tool-augmented reasoning &
Real-world multi-day travel itinerary planning covering transportation, accommodation, dining, attractions, and inter-city routing under constraints &
Delivery rate; commonsense pass rate; hard-constraint pass rate; final pass rate
\\
\hline

\textbf{TP-RAG}~\cite{ni-etal-2025-tp} &
Language + Geovector + Trajectory &
Spatiotemporal-aware travel planning; retrieval-augmented reasoning; trajectory-level knowledge utilization; temporal adaptability &
Single-city multi-day travel planning with query-dependent POIs and retrieved tourist trajectories; spatially efficient, temporally coherent, and query-relevant itineraries generation &
Commonsense + spatial + temporal + POI semantic + query relevance metrics \\
\hline

\end{tabular}
\caption{\rev{Summary of existing UFMs benchmarks by data modalities involved, evaluated abilities, representative tasks, and evaluation metrics.}}
\label{tab:urban_benchmarks}
\end{table*}

\subsection{Benchmarks}
Recent years have witnessed a rapid emergence of benchmarks designed to evaluate UFMs across heterogeneous data modalities, reasoning paradigms, and application scenarios. As summarized in Table~\ref{tab:urban_benchmarks}, existing benchmarks can be broadly grouped into three themes: spatiotemporal reasoning and decision-making, urban perception and embodiment, and specialized urban task. 

Spatiotemporal reasoning and decision-making benchmarks focus on assessing whether large models can reason over dynamic urban processes involving space, time, and actions. Benchmarks such as USTBench~\cite{lai2025ustbench}, CityBench~\cite{feng2024citybench}, STBench~\cite{li2025stbench}, and STARK~\cite{quan2025benchmarking} evaluate models on tasks ranging from mobility and congestion forecasting to localization, trajectory reasoning, and navigation-style inference. A key distinction among them lies in how reasoning is evaluated. CityBench, STBench, and STARK primarily emphasize task-level correctness, measuring whether models can infer spatial–temporal relations or numerical outcomes under increasing complexity. In contrast, USTBench explicitly introduces process-level evaluation, probing intermediate reasoning steps (e.g., understanding, forecasting, planning, and reflection) rather than only final outcomes. Empirical findings across these benchmarks consistently show that while LLMs can handle short-horizon prediction and symbolic reasoning, they struggle with long-horizon planning, geometric reasoning, and error correction under feedback, indicating a gap between linguistic reasoning and actionable urban intelligence.

Urban perception and embodiment benchmarks extend evaluation into perceptual and embodied urban settings. Benchmarks such as UrBench~\cite{zhou2025urbench}, CityEQA~\cite{zhao-etal-2025-cityeqa}, and UrbanVideo-Bench~\cite{zhao-etal-2025-urbanvideo} examine how multimodal models interpret urban scenes, reason over visual evidence, and support navigation. UrBench focuses on multi-view urban perception and cross-view consistency, evaluating large multimodal models on geo-localization, scene understanding and reasoning, object understanding, and consistency across street-view, aerial-view, and satellite-view imagery through diverse QA tasks. CityEQA and UrbanVideo-Bench move beyond static perception toward embodied and temporal understanding: CityEQA evaluates embodied active exploration and long-horizon spatial reasoning by requiring agents to navigate a realistic urban simulator to answer open-vocabulary questions, while UrbanVideo-Bench assesses urban video understanding across recall, perception, reasoning, and navigation using motion-rich city-scale video sequences. Collectively, results from these benchmarks reveal that although vision–language models can capture local visual cues, they still struggle with persistent spatial memory, causal and relational reasoning over time, and long-horizon consistency, particularly in open-ended, city-scale embodied settings that require integrating perception, memory, and decision-making.

Specialized urban task benchmarks target professional urban domain knowledge rather than perception or control. 
CityLens~\cite{liu2025citylens} highlights the difficulty of inferring latent socioeconomic indicators from urban imagery, showing that visual grounding alone is insufficient for reliable quantitative prediction. 
LLM-UrbanPlan~\cite{zhao2025urban} evaluates LLMs on hundreds of real-world urban planning tasks, spanning planning documentation, professional examinations, routine data analysis, algorithmic support, and academic writing, with a combination of automatic metrics and expert human judgments. 
TravelPlanner~\cite{xie2024travelplanner} assesses long-horizon, constraint-intensive planning, requiring models to generate realistic multi-day travel itineraries under hard constraints on time, budget, transportation, and logistics, and evaluating plans using fine-grained constraint-satisfaction and execution-validity metrics.
TP-RAG~\cite{ni-etal-2025-tp} evaluates retrieval-augmented travel planning with query-specific spatiotemporal contexts and trajectory-level external knowledge, examining whether models can generate itineraries that are spatially coherent, temporally adaptive, and semantically aligned with user constraints. 
These benchmarks expose a remarkable limitation of current foundation models: despite strong general language fluency and visual–semantic alignment, they struggle to support reliable urban analysis and planning, showing weak causal grounding for quantitative inference and uneven mastery of domain-specific constraints and standards. As a result, they often produce outputs that appear plausible but are brittle, shallow, or incorrect when faced with professional urban tasks. Even advanced reasoning-oriented models fall short of professional-level competence, underscoring the challenge of transferring general-purpose foundation models into specialized urban domains.

\subsection{Datasets}

\begin{table}[th]
\centering
\tiny
\begin{tblr}{
  colspec = { 
    Q[m, l, 1cm]
    Q[m, l, 2.4cm]
    Q[m, l, 5cm]
    Q[m, l, 3.5cm]
  },
  hlines,       
  vline{2-4} = {solid}, 
  rows = {m},   
  row{1} = {font=\bfseries}, 
}
Modality & Data Types & Datasets & Data Sources \\

\SetCell[r=4]{c} \textbf{Language} 
  & Urban planning texts \& QA
  & \href{https://github.com/tsinghua-fib-lab/PlanBench}{PlanBench}~\cite{zheng2025urbanplanbench}; \href{https://github.com/JIANGYUE61610306/UrbanLLM/tree/main/Urbanllm_data}{UrbanLLM-Data}~\cite{jiang2024urbanllm} 
  & Urban Planning Textbooks; \href{https://data.gov.sg/}{Singapore Open Data} \\

  & Travel QA
  & \href{https://huggingface.co/datasets/osunlp/TravelPlanner}{TravelPlanner}~\cite{xie2024travelplanner}; \href{https://huggingface.co/datasets/soniawmeyer/reddit-travel-QA-finetuning}{Travel-QA}~\cite{meyer2024comparison}; \href{https://github.com/hsaest/QUERT/tree/main}{QUERT-Data}~\cite{xie2023quert} 
  & \href{https://developers.google.com/maps/}{Google Maps}; \href{https://developers.reddit.com/}{Reddit}; \href{https://open.fliggy.com/}{Fliggy} \\

  & Geoscience knowledge QA
  & \href{https://github.com/davendw49/k2}{GeoBench, GeoSignal}~\cite{deng2024k2}; \href{https://github.com/AGI-GIS/BB-GeoGPT}{BB-GeoGPT-Data}~\cite{zhang2024bb} 
  & Geoscience Literatures., Wikipedia. \\

  & Environmental event QA 
  & \href{https://github.com/SusuXu-s-Lab/Hierarchical-Earthquake-Casualty-Information-Retrieval}{Global Earthquake}~\cite{wang2024near}; \href{https://github.com/tdiggelm/climate-fever-dataset}{Climate Fever}~\cite{diggelmann2020climate}; \href{https://huggingface.co/datasets/mbzuai-oryx/Clima500}{Climate500}~\cite{mullappilly2023arabic} 
  & \href{https://earthquake.usgs.gov/data/pager/}{PAGER system}; N.A.; \href{https://github.com/SmartDataAnalytics/Climate-Bot}{CCMRC} \\

\SetCell[r=7]{c} \textbf{Vision} 
  & \SetCell[r=5]{l} Street-view imagery 
    & \href{https://www.nuscenes.org/}{nuScenes}~\cite{caesar2020nuscenes}; \href{https://www.nuscenes.org/nuplan}{nuPlan}~\cite{nuplan} 
    & \href{https://www.nuscenes.org/}{nuScenes} \\
  
  & & \href{https://github.com/tsinghua-fib-lab/UrbanKG-KnowCL}{Know-CL-Data}~\cite{liu2023knowledge} 
    & \href{https://lbsyun.baidu.com/}{Baidu Maps}, \href{https://www.google.com/streetview/}{Google Street} \\
  
  & & \href{https://waymo.com/open}{WOMD}~\cite{ettinger2021large}; \href{https://uwaterloo.ca/watonobus/downloads}{WATonoBus}~\cite{bhatt2024watonobus}; \href{https://github.com/microsoft/AirSim}{AirSim}~\cite{shah2017airsim}
    & \href{https://waymo.com/open}{Waymo}; \href{https://uwaterloo.ca/watonobus/}{WATonoBus Proj.}; \href{https://www.microsoft.com/en-sg}{Microsoft} \\
  
  & & \href{https://github.com/carla-simulator/carla}{CARLA}~\cite{dosovitskiy2017carla}; \href{https://github.com/llmbev/talk2bev?tab=readme-ov-file}{Talk2BEV}~\cite{choudhary2024talk2bev} 
    & N.A. \\
  
  & & \href{https://github.com/JinkyuKimUCB/BDD-X-dataset}{BDD-X}~\cite{kim2018textual}; \href{https://github.com/xmed-lab/NuInstruct}{NuInstruct}~\cite{ding2024holistic}; \href{https://github.com/OpenDriveLab/DriveLM}{DriveLM}~\cite{sima2024drivelm} 
    & N.A. \\

  & Remote sensing imagery
  & \href{https://github.com/tsinghua-fib-lab/UrbanKG-KnowCL}{Know-CL-Data}~\cite{liu2023knowledge}; \href{https://captain-whu.github.io/DiRS/}{Million-AID}~\cite{long2021creating} 
  & \href{https://www.arcgis.com/}{ArcGIS}; \href{https://earth.google.com/web/}{Google Earth} \\

  & Meteorological raster
  & \href{https://cds.climate.copernicus.eu/datasets/reanalysis-era5-single-levels?tab=overview}{ERA5}~\cite{hersbach2020era5}; \href{https://extremeweatherdataset.github.io/}{ExtremeWeather}~\cite{racah2017extremeweather} 
  & \href{https://www.ecmwf.int/}{ECMWF}; N.A. \\

\SetCell[r=13]{c} \textbf{Time Series} 
  & Energy consumption 
  & \href{https://github.com/zhouhaoyi/ETDataset}{ETT}~\cite{zhou2021informer}; \href{https://github.com/laiguokun/multivariate-time-series-data}{Electricity}~\cite{lai2018modeling}; \href{https://github.com/laiguokun/multivariate-time-series-data}{Solar-Energy}~\cite{lai2018modeling} 
  & N.A.; \href{https://archive.ics.uci.edu/dataset/321/electricityloaddiagrams20112014}{UCI}; \href{https://www.nrel.gov/grid/solar-power-data}{Nrel} \\

  & \SetCell[r=7]{l} Traffic flow \& speed 
    & \href{https://github.com/laiguokun/multivariate-time-series-data}{Traffic}~\cite{lai2018modeling}; \href{https://github.com/Davidham3/STSGCN}{PEMS}~\cite{song2020spatial}; \href{https://github.com/VeritasYin/STGCN_IJCAI-18}{PEMSD7(M/L)}~\cite{yu2018spatio}; \href{https://github.com/liuxu77/LargeST}{LargeST}~\cite{liu2023largest} 
    & \href{https://pems.dot.ca.gov/}{Caltrans PEMS} \\
  
  & & \href{https://github.com/liyaguang/DCRNN}{PEMS-BAY, METR-LA}~\cite{li2018diffusion} 
    & \href{https://pems.dot.ca.gov/}{Caltrans PEMS}, \href{https://www.metro.net/}{Metro} \\

  & & \href{https://github.com/RobinLu1209/ST-GFSL?tab=readme-ov-file}{Chengdu/Shenzhen-didi}~\cite{lu2022spatio}; \href{https://github.com/JingqingZ/BaiduTraffic}{Q-Traffic}~\cite{liao2018deep}
    & \href{https://web.didiglobal.com/}{DiDi}; \href{https://lbsyun.baidu.com/}{Baidu Maps} \\
  
  & & \href{https://github.com/tsinghua-fib-lab/UniST/tree/main/dataset}{Traffic(HZ/CD/JN/NJ)}~\cite{yuan2024unist};  \href{https://github.com/lehaifeng/T-GCN/tree/master/data}{SZ-TAXI}~\cite{zhao2019t} 
    & N.A. \\
  
  & & \href{https://github.com/zhiyongc/Seattle-Loop-Data?tab=readme-ov-file}{Loop-Seattle}~\cite{cui2018deep}; \href{https://github.com/RomainLITUD/Multistep-Traffic-Forecasting-by-Dynamic-Graph-Convolution}{ROTTERDAM}~\cite{li2021multistep}; \href{https://github.com/TolicWang/DeepST/tree/master/data/TaxiBJ}{TaxiBJ}~\cite{zhang2017deep} 
    & N.A. \\
  
  & & \href{https://github.com/JinleiZhangBJTU/ResNet-LSTM-GCN}{BJ-SW}~\cite{zhang2020deep}; \href{https://github.com/ivechan/PVCGN}{HZ, SHMETRO}~\cite{liu2020physical} 
    & N.A. \\
  
  & & \href{https://huggingface.co/datasets/bjdwh/UrbanGPT_ori_stdata/tree/main}{NYC/CHI-TAXI}~\cite{li2024urbangpt}; \href{https://github.com/KL4805/CrossTReS}{NYC/CHI/DC-BIKE}~\cite{jin2022selective} 
    & \href{https://opendata.cityofnewyork.us/}{NYC OpenData}, \href{https://data.cityofchicago.org/}{CHI Portal}, \href{https://capitalbikeshare.com/system-data}{CapitalBike} \\

  & \SetCell[r=3]{l} Weather conditions 
    & \href{https://github.com/zhouhaoyi/Informer2020}{Weather}~\cite{zhou2021informer}; \href{https://github.com/cruiseresearchgroup/PISA-PromptCast}{City Temp.}~\cite{xue2023promptcast} 
    & \href{https://www.ncei.noaa.gov/data/local-climatological-data/}{NCEI}; \href{https://udayton.edu/}{Univ. of Dayton} \\
  
  & & \href{https://github.com/Echohhhhhh/GSNet}{NYC/CHI-Weather}~\cite{wang2021gsnet} 
    & \href{https://opendata.cityofnewyork.us/}{NYC OpenData}, \href{https://data.cityofchicago.org/}{CHI Portal} \\
  
  & & \href{https://github.com/BruceBinBoxing/Deep_Learning_Weather_Forecasting}{WD\_BJ}~\cite{wang2019deep}; \href{https://github.com/mb-Ma/HiSTGNN}{WD\_USA, WD\_ISR}~\cite{ma2023histgnn} 
    & \href{https://github.com/AIChallenger/AI_Challenger_2018}{AI Challenger}; \href{https://www.kaggle.com/datasets/selfishgene/historical-hourly-weather-data/data}{Kaggle} \\

  & Air quality
  & \href{https://www.microsoft.com/en-us/research/project/urban-air/}{UrbanAir}~\cite{zheng2015forecasting}; \href{https://github.com/shuowang-ai/PM2.5-GNN}{KnowAir}~\cite{wang2020pm2}
  & \href{https://www.microsoft.com/en-us/research/lab/microsoft-research-asia/}{MSRA}; \href{https://english.mee.gov.cn/}{MEE} \\

  & Crime 
  & \href{https://github.com/LZH-YS1998/STHSL}{NYC/CHI-Crime}~\cite{li2022spatial} 
  & \href{https://opendata.cityofnewyork.us/}{NYC OpenData}, \href{https://data.cityofchicago.org/}{CHI Portal} \\

\SetCell[r=3]{c} \textbf{Trajectory} 
  & POI check-ins 
  & \href{https://sites.google.com/site/yangdingqi/home/foursquare-dataset}{Foursquare-NYC/Tokyo}~\cite{yang2014modeling}; \href{https://snap.stanford.edu/data/loc-gowalla.html}{Gowalla}~\cite{cho2011friendship}; \href{https://business.yelp.com/data/resources/open-dataset/}{Yelp}~\cite{wang2019kgat} 
  & \href{https://foursquare.com/}{Foursquare}; N.A.; \href{https://www.yelp.com/}{Yelp} \\
 
  & \SetCell[r=2]{l} GPS trajectories 
  & \href{https://www.microsoft.com/en-us/research/publication/geolife-gps-trajectory-dataset-user-guide/}{GeoLife}~\cite{zheng2010geolife}; \href{https://www.microsoft.com/en-us/research/publication/t-drive-trajectory-data-sample/}{T-Drive}~\cite{yuan2010t}
  & \href{https://www.microsoft.com/en-us/research/lab/microsoft-research-asia/}{MSRA}\\
  & & \href{https://archive.ics.uci.edu/dataset/339/taxi+service+trajectory+prediction+challenge+ecml+pkdd+2015}{Portal-Taxi}~\cite{jiang2023continuous} & N.A.\\

\SetCell[r=3]{c} \textbf{Geovector} 
  & POI data 
  & \href{https://download.geofabrik.de/}{OSM Data Extracts}~\cite{tempelmeier2021geovectors} 
  & \href{https://www.openstreetmap.org/}{OpenStreetMap} \\

  & Road networks 
  & \href{https://github.com/tsinghua-fib-lab/DRL-urban-planning}{HLG, DHM}~\cite{zheng2023spatial}; \href{https://github.com/changyanchuan/SARN}{SARN-Data}~\cite{chang2023spatial} 
  & \href{https://www.openstreetmap.org/}{OpenStreetMap} \\

  & Urban regions 
  & \href{https://github.com/tsinghua-fib-lab/DRL-urban-planning}{HLG, DHM}~\cite{zheng2023spatial}; \href{https://github.com/jackmiemie/GMEL/tree/master/data/PLUTO}{PLUTO}~\cite{liu2020learning} 
  & \href{https://www.openstreetmap.org/}{OpenStreetMap}; \href{https://opendata.cityofnewyork.us/}{NYC OpenData} \\

\SetCell[r=6]{c} \textbf{Multimodal} 

  & Urban map images with textual queries
  & \href{https://github.com/zhuchichi56/PlanGPT-VL?tab=readme-ov-file}{PlanBench-V}~\cite{zhu-etal-2025-plangpt-vl}
  & Urban Planning Bureaus \\

  & Travel queries with POIs \& mobility trajectories
  & \href{https://github.com/usail-hkust/TP-RAG}{TP-RAG}~\cite{ni-etal-2025-tp}
  & \href{https://lbsyun.baidu.com/}{Baidu Maps}, \href{https://www.baidu.com/}{Baidu Search} \\

  & Street-view images with geo-localization queries
  & \href{https://github.com/yeyimilk/LLMGeo}{LLMGeo-Data}~\cite{wang2024llmgeo}
  & \href{https://www.google.com/streetview/}{Google Street} \\

  & Weather imagery with textual QA
  & \href{https://github.com/chengqianma/WeatherQA}{WeatherQA}~\cite{ma2024weatherqa} 
  & \href{https://www.spc.noaa.gov/misc/about.html\#Mesoscale Discussions}{SPC} \\

  & Satellite imagery with POIs
  & \href{https://github.com/axin1301/satellite-imagery-POI}{Beijing}~\cite{xi2022beyond}; \href{https://github.com/bailubin/MMGR}{Shanghai}~\cite{bai2023geographic} 
  & \href{https://lbs.qq.com/getPoint/}{Tencent Maps}; \href{https://lbs.amap.com/}{AMap} \\

  & Satellite imagery with POIs \& human mobility data
  & \href{https://github.com/porterjenkins/region-encoder}{NYC, CHI}~\cite{jenkins2019unsupervised} 
  & \href{https://developers.google.com/maps/}{Google Maps}, \href{https://opendata.cityofnewyork.us/}{NYC OpenData}, \href{https://data.cityofchicago.org/}{CHI Portal}, \href{https://foursquare.com/}{Foursquare} \\

\textbf{Other} 
  & Traffic signal 
  & \href{https://github.com/wingsweihua/colight/tree/master/data}{JN/HZ/NYC-SN}~\cite{wei2019colight}; \href{https://github.com/Chacha-Chen/MPLight}{Manhattan-SN}~\cite{chen2020toward} 
  & \href{https://www.nyc.gov/html/dot/html/infrastructure/signals.shtml}{NYC Sig}; \href{https://www.openstreetmap.org/}{OpenStreetMap} \\

\end{tblr}

\caption{\rev{Summary of datasets that could be used for building UFMs.}}
\label{tab:ufm_datasets}
\end{table}

Building UFMs relies on a diverse ecosystem of datasets spanning language, vision, time series, trajectories, geovectors, and multimodal data. As summarized in Table~\ref{tab:ufm_datasets}, existing datasets can be systematically organized by modality and data type, reflecting both the heterogeneous nature of urban data and the different functional roles these sources might play in UFMs’ pre-training, adaptation, and evaluation.

Language-based datasets serve as the primary carriers of high-level knowledge for UFMs, encompassing not only semantic understanding but also procedural logic and normative constraints that underpin urban analysis and planning tasks. Within this modality, urban planning texts and QA datasets such as PlanBench~\cite{zheng2025urbanplanbench} and UrbanLLM-Data~\cite{jiang2024urbanllm} curate professional documents, regulations, and expert-level questions that encode planning standards and domain-specific terminology. Travel-oriented QA datasets, including TravelPlanner~\cite{xie2024travelplanner}, Travel-QA~\cite{meyer2024comparison}, and QUERT-Data~\cite{xie2023quert}, focus on itinerary construction, constraint satisfaction, and commonsense reasoning under real-world travel constraints. Geoscience and environmental knowledge is captured by QA-style datasets such as GeoBench~\cite{deng2024k2}, GeoSignal~\cite{deng2024k2}, and BB-GeoGPT-Data~\cite{zhang2024bb}, while event-centric datasets like Global Earthquake~\cite{wang2024near}, Climate-Fever~\cite{diggelmann2020climate}, and Climate500~\cite{mullappilly2023arabic} emphasize factual verification and causal explanation of climate- and disaster-related claims. These datasets equip UFMs with the ability to interpret expert discourse, adhere to regulatory and commonsense constraints, and perform reasoning that aligns with real-world urban policies, environmental dynamics, and societal norms.

Vision-based datasets supply UFMs with direct sensory evidence of the physical city, capturing visual structures and environmental conditions from street-level scenes to aerial and atmospheric observations. Street-view and autonomous-driving datasets such as nuScenes~\cite{caesar2020nuscenes}, nuPlan~\cite{nuplan}, WOMD~\cite{ettinger2021large}, WATonoBus~\cite{bhatt2024watonobus}, CARLA~\cite{dosovitskiy2017carla}, AirSim~\cite{shah2017airsim}, BDD-X~\cite{kim2018textual}, NuInstruct~\cite{ding2024holistic}, DriveLM~\cite{sima2024drivelm}, and Talk2BEV~\cite{choudhary2024talk2bev} provide multi-view imagery, sensor fusion data, and increasingly language-annotated reasoning signals for perception, localization, and decision-making. Remote sensing datasets such as Million-AID~\cite{long2021creating} and Know-CL-Data~\cite{liu2023knowledge} support large-scale land-use analysis and urban morphology understanding, while meteorological raster products including ERA5~\cite{hersbach2020era5} and ExtremeWeather~\cite{racah2017extremeweather} enable coupling between atmospheric processes and urban dynamics.
Through learning from these visually grounded data sources, UFMs can internalize fine-grained spatial layouts, multi-scale urban form, and the coupling between built environments and natural processes—capabilities that cannot be reliably inferred from textual or abstract representations alone.

Time-series datasets capture the continuous temporal signals of urban systems, providing essential observations of how infrastructure, mobility, environment, and social activities evolve over time. 
Energy consumption datasets such as ETT~\cite{zhou2021informer}, Electricity~\cite{lai2018modeling}, and Solar-Energy~\cite{lai2018modeling} support load forecasting and infrastructure analysis. Traffic flow and speed datasets form the largest and most mature category, including Traffic~\cite{lai2018modeling}, PEMS~\cite{song2020spatial}, PEMS-D7(M/L)~\cite{yu2018spatio}, METR-LA~\cite{li2018diffusion}, LargeST~\cite{liu2023largest}, Q-Traffic~\cite{liao2018deep}, SZ-TAXI~\cite{zhao2019t}, TaxiBJ~\cite{zhang2017deep}, and many city-specific benchmarks.
These traffic datasets are primarily used for forecasting congestion, estimating travel time, and learning spatiotemporal dependencies in large-scale transportation networks.
Weather condition datasets, including Weather~\cite{zhou2021informer}, City Temperature~\cite{xue2023promptcast}, and NYC/CHI-Weather~\cite{wang2021gsnet}, air-quality datasets such as UrbanAir~\cite{zheng2015forecasting} and KnowAir~\cite{wang2020pm2}, and crime time series~\cite{li2022spatial} further extend coverage to environmental and social signals. 
By learning from these diverse temporal streams, UFMs can develop robust representations of periodicity, long-range temporal dependencies, and cross-domain spatiotemporal interactions that underpin forecasting, monitoring, and adaptive urban decision-making.

Trajectory datasets record the movement of people and vehicles through urban space, offering direct evidence of how individuals and populations interact with the built environment over time. POI check-in datasets such as Foursquare~\cite{yang2014modeling}, Gowalla~\cite{cho2011friendship}, and Yelp~\cite{wang2019kgat} encode semantic activity preferences and are widely used for next-POI prediction and recommendation tasks. GPS trajectory datasets, including GeoLife~\cite{zheng2010geolife}, T-Drive~\cite{yuan2010t}, and Portal-Taxi~\cite{jiang2023continuous}, provide fine-grained spatial–temporal traces for modeling route choice, origin–destination flows, and urban movement dynamics.
Through training on such mobility traces, UFMs can acquire an understanding of habitual travel behaviors, spatial constraints, and activity semantics, enabling more realistic modeling of urban dynamics and human-centric decision processes.

Geovector datasets explicitly encode the fixed spatial structure of cities by representing places, roads, and parcels as vectorized geometric and semantic entities. 
POI datasets and road network derived from OpenStreetMap~\cite{tempelmeier2021geovectors,chang2023spatial,zheng2023spatial} are commonly used to construct spatial graphs and semantic maps.
More specialized datasets, including HLG and DHM~\cite{zheng2023spatial}, SARN-Data~\cite{chang2023spatial}, and PLUTO~\cite{liu2020learning}, provide curated representations of road hierarchies, land use, and administrative regions. Such structured spatial annotations are valuable for UFMs, as they enable models to learn hierarchical spatial semantics, functional zoning, and cross-scale relationships that are difficult to infer from other data mobilities, thereby strengthening models' spatial reasoning and urban-context understanding.

Multimodal datasets provide a unified learning substrate by systematically coupling heterogeneous urban data sources, such as text, maps, imagery, trajectories, and sensor signals, thereby exposing UFMs to the interactions and dependencies across modalities. 
Representative datasets include PlanBench-V~\cite{zhu-etal-2025-plangpt-vl} for map-based planning, TP-RAG~\cite{ni-etal-2025-tp} for travel planning with retrieval over POIs and trajectories, LLMGeo~\cite{wang2024llmgeo} for street-view geo-localization, WeatherQA~\cite{ma2024weatherqa} for meteorological question answering, and satellite imagery combined with POIs and human mobility data for regional representation learning~\cite{xi2022beyond,bai2023geographic,jenkins2019unsupervised}.
By learning from such aligned multimodal inputs, UFMs can move beyond isolated perception or language understanding toward holistic urban intelligence, supporting coordinated reasoning, long-horizon planning, and informed decision-making in complex urban environments.

}

\section{Applications} \label{application}
In this section, we discuss several potential urban scenarios that can benefit from UFMs. Although extensive efforts have already been made for these areas by using AI technologies, we envision UFMs will enable more comprehensive solutions for each application domain.

\subsection{Transportation}
Intelligent transportation systems have profoundly revolutionized various aspects of our society, ranging from road safety~\cite{wang2023transportation} and public transportation~\cite{liu2020polestar} to traffic management decision-making~\cite{lai2023large}. Currently, an immense volume of data reflecting traffic dynamics has been generated and collected through a variety of sources, including social media feeds, digital government platforms, cameras, and massive IoT devices. Based on these data, a few studies have attempted to develop foundation models for transportation, such as TransGPT~\cite{wang2024transgpt}, TrafficGPT~\cite{zhang2023trafficgpt}, and TengYun~\cite{zhao2023parallel}. Nevertheless, these works primarily focus on specific aspects of transportation, \eg~text-related tasks. Developing versatile UFMs capable of addressing a diverse set of transportation tasks still remains an open problem. One major challenge lies in that data from transportation systems is inherently heterogeneous and multi-modal, which encompasses time series, trajectories, images, text, \etc 
Therefore, UFMs need to effectively handle the heterogeneity and multi-modality of transportation data and maintain satisfactory performance across diverse applications.

\subsection{Urban Planning}
Effective urban planning is crucial in developing intelligent, efficient, and eco-friendly cities. The widely available urban data, such as road network structure and human mobility, has indeed uncovered the underlying issues of a city, providing urban planners with valuable insights for making informed decisions~\cite{zheng2014urban}. UFMs can bring substantial benefits to urban planning. On the one hand, they can assist in analyzing vast amounts of urban data, generating creative ideas to support urban planners and policymakers for better future planning formulation~\cite{wang2023towards}. On the other hand, UFMs can also enhance the participatory planning process by mining public feedback from various digital platforms~\cite{zhou2024large}. For instance, citizens can have real-time interactions with UFMs to provide suggestions about urban infrastructures, public transportation schedules, and health services. This results in a significant shift towards more inclusive and responsive urban planning, where voice of the public plays an important role in shaping urban landscapes.

\subsection{Energy Management}
AI techniques have proven to be beneficial for the operational procedures of urban energy systems~\cite{zhan2022deepthermal,yao2023machine,zhang2021intelligent,zhang2022multi}. In addition, the proliferation of urban energy data makes it possible to build domain-specific foundation models. Thus, the application of UFMs in the domain of energy management holds great promise. 
As one of the initial efforts in this area, Huang et al.~\cite{huang2023large} explore a set of capabilities of existing foundation models (\ie GPT-4 and GPT-4 Vision), showcasing promising results in addressing intricate challenges in power systems.
However, a potential challenge is that most energy data, such as household electricity consumption, naturally possess high sensitivity and personal privacy information, making the relevant departments unwilling to disclose the trained models and related data to the public. Privacy preservation techniques like federated learning~\cite{yang2019federated} might be promising solutions to fully unleash the potential of foundation models in energy management.

\subsection{Environmental Monitoring}
The rapid urbanization progress has generated numerous environmental issues, such as air pollution, water contamination, and resource depletion, posing tremendous threats to public health and cities' sustainable development. Recently, Vaghefi et al.~\cite{vaghefi2023chatclimate} introduced ChatClimate, a retrieval-augmented framework to enhance the professional skills of LLMs in the field of climate change by incorporating up-to-date domain-specific data. Furthermore, a series of research studies~\cite{nguyen2023climax,bodnar2024aurora,bi2023accurate,lam2023learning} have developed specialized foundation models for climate-related tasks, using numerical data from reanalysis databases, satellites, and in-situ sensors. Despite these advancements, the utilization of UFMs for broader environmental problems, such as air pollution analysis~\cite{han2023machine}, still remains under-explored. The versatile UFMs, potentially integrated with powerful pre-existing foundation models, could significantly enrich existing research and further facilitate a wide range of environmental applications.

\subsection{Public Safety and Security}
Public safety and security is a critical aspect of urban life, directly impacting the well-being of citizens. Building foundation models for this area can advance various applications, including crime prevention, emergency response, and disaster management. Recently, some LLM-based approaches have been explored for public safety and security, such as TrafficSafetyGPT~\cite{zheng2023trafficsafetygpt} and ChatLaw~\cite{cui2023chatlaw}. These models fine-tuned on instructions derived from human annotators and ChatGPT, demonstrate great promise in producing responses with reliable traffic safety and legal domain knowledge. Currently, the full potential of LLMs in enhancing urban public safety remains largely untapped. For example, LLMs might analyze crime reports to discern emerging patterns or trends, helping law enforcement agencies identify potential hotspots of criminal activities and allocate resources more effectively.
To achieve this goal, future research might focus on integrating external tools, such as search engines and crime prediction models, to improve the problem-solving capabilities of LLMs.

\subsection{Other Domains}
UFMs have the potential to drive transformative changes in other related domains. In economy, UFMs can support financial decision-making at the urban scale by analyzing economic data to predict market trends, assess investment opportunities, and optimize budget allocation for urban development projects~\cite{li2023large}. For example, UFMs could analyze real estate trends, urban economic activity, and public spending patterns to generate insights that inform the design of more financially sustainable urban policies. In healthcare, UFMs can help optimize healthcare resource distribution by analyzing historical usage patterns and predicting future demands for medical infrastructure~\cite{he2023survey,moor2023foundation}, \eg hospitals, emergency services, and vaccination centers. Furthermore, by unlocking the emergent abilities, UFMs could be prompted with minimal human instructions to dynamically define fundamentally new urban tasks, such as creating actionable strategies to improve urban well-being, thereby further advancing the development of intelligent cities.

\section{Conclusion and Future Work} \label{conclusion}

\rev{This paper presented a comprehensive review of Urban Foundation Models (UFMs), emphasizing their transformative potential for advancing the understanding of urban dynamics and significantly enhancing Urban General Intelligence (UGI). We began by defining the concepts of UGI and UFMs, and highlighting the key challenges associated with UFMs' development. Subsequently, we categorized existing UFM-related studies through a data-centric taxonomy framework, structured around urban data modalities and types, to provide a coherent perspective on ongoing research. Furthermore, we proposed forward-looking solutions aimed at addressing these challenges, paving the way for more versatile and effective UFMs. 
Additionally, we systematically summarized and discussed existing benchmarks and datasets related to UFMs. 
Finally, we explored the potential applications and positive impacts of UFMs across various critical urban domains.
We hope this review serves as a valuable resource for researchers and practitioners, fostering further exploration and innovation in this emerging field.}

Looking ahead, the evolution of UFMs holds significant potential for groundbreaking advancements. Future research is expected to prioritize the integration and analysis of multi-source, multi-granularity, and multimodal urban data to support an even broader spectrum of urban applications and scenarios. Another key focus will be on developing versatile UFMs capable of real-time urban data analysis to deliver timely and actionable urban insights. Strengthening their spatio-temporal reasoning abilities will further enhance their performance in dynamic urban environments.
Additionally, ensuring the ethical development of UGI will require a careful balance between leveraging urban data and safeguarding privacy and security. Establishing equitable compensation mechanisms for data sharing and model development will also be critical to fostering sustainable and collaborative progress in UFM innovation. We are enthusiastic about the future of UFMs and their potential to enable smarter, more resilient, and adaptive urban spaces.

\begin{acks}
This work was supported in part by the National Key R\&D Program of China (Grant No.2023YFF0725-001),in part by the National Natural Science Foundation of China (Grant No.62572417, No.92370204),in part by the guangdong Basic and Applied Basic Research Foundation (Grant No.2023B1515120057),in part by the Key-Area Special Project of Guangdong Provincial Ordinary Universities (2024ZDZX1007).
\end{acks}

\bibliography{refs}
\bibliographystyle{ACM-Reference-Format}

\end{document}